\documentclass[12pt]{article}
\usepackage{amsmath,amsthm}
\usepackage{eufrak}
\usepackage[pdftex]{graphicx}

\newcommand{\ar}{\renewcommand{\arraystretch}{1}} 
\DeclareMathAlphabet{\bb}{U}{msb}{m}{n} \gdef\C{\bb C} \gdef\dZ{\bb
Z}   \gdef\dS{\bb S} \gdef\R{\bb R}
\gdef\K{\bb K} \gdef\BH{\bb H} \gdef\F{\bb F} 

 \DeclareMathOperator{\spin}{{\bf
Spin}} 
\DeclareMathOperator{\fD}{\mathfrak{D}}

\DeclareMathOperator{\Sym}{Sym}

 \DeclareMathOperator{\SL}{SL}
\DeclareMathOperator{\SO}{SO}\DeclareMathOperator{\SU}{SU}

\newcommand{\cA}{\mathcal{A}}

\newcommand{\sA}{{\sf A}}
\newcommand{\sB}{{\sf B}}

\newcommand{\sH}{{\sf H}}

\newcommand{\sX}{{\sf X}}
\newcommand{\sY}{{\sf Y}}

\newcommand{\bsH}{{\boldsymbol{\sf H}}}

\newcommand{\bJ}{{\bf J}}

\newcommand{\fA}{\mathfrak{A}}

\newcommand{\fg}{\mathfrak{g}}

\newcommand{\cl}{C\kern -0.2em \ell}

\newcommand{\ld}{\left[}
\newcommand{\rd}{\right]}

\begin{document}
\title{Lorentz group and mass spectrum of elementary particles}
\author{V.~V. Varlamov\thanks{Siberian State Industrial University,
Kirova 42, Novokuznetsk 654007, Russia, e-mail:
varlamov@sibsiu.ru}}
\date{}
\maketitle
\begin{abstract}
Mass spectrum of localized states (elementary particles) of single quantum system is studied in the framework of Heisenberg's scheme. Localized states are understood as cyclic representations of a group of fundamental symmetry (Lorentz group) within a Gelfand-Neumark-Segal construction. It is shown that state masses of lepton (except the neutrino) and hadron sectors of matter spectrum are proportional to the rest mass of electron with an accuracy of $0,41\%$.
\end{abstract}

\section{Introduction}
One of the most important undecided problems in theoretical physics is a description of mass spectrum of elementary particles (it is one from 30 problems in Ginzburg's list \cite{Ginz}). It is well known that a quark model, based on the flavor $\SU(3)$ group, does not explain the mass spectrum of elementary particles. The Gell-Mann--Okubo mass formula explains only a mass splitting within supermultiplets of $\SU(3)$-theory, namely, a hypercharge mass splitting within supermultiplets and a charge splitting within isotopic multiplets belonging to a given supermultiplet. An action of the group $\SU(3)$ is analogous to Zeeman effect in atomic spectra, that is, this action leads to different mass levels within charge multiplets via $\SU(3)/\SU(2)$-reduction. Masses of particles, belonging to a given supermultiplet of $\SU(3)$ in $\SU(3)/\SU(2)$-reduction, are defined by the Gell-Mann--Okubo mass formula \cite{Gel61,OR64} (quadratic case)
\begin{equation}\label{GMOQ}
m^2=m^2_0+\alpha+\beta Y+\gamma\left[I(I+1)-\frac{1}{4}Y^2\right]
+\alpha^\prime-\beta^\prime
Q+\gamma^\prime\left[U(U+1)-\frac{1}{4}Q^2\right],
\end{equation}
where $Q$ and $Y$ are charge and hypercharge of the particle, $I$ and $U$ are isotopic spins. The coefficients $\alpha$, $\alpha^\prime$, $\beta$, $\beta^\prime$, $\gamma$, $\gamma^\prime$ are satisfy the relation
\[
\frac{\alpha^\prime}{\alpha}=\frac{\beta^\prime}{\beta}=\frac{\gamma^\prime}{\gamma}=\theta,\quad
|\theta|\ll 1.
\]
In the case when the condition $\delta m^2/m^2_0\ll 1$ is fulfilled, the
quadratic mass formula (\ref{GMOQ}) can be replaced by the linear
mass formula
\begin{equation}\label{GMOL}
m=m_0+\alpha+\beta Y+\gamma\left[I(I+1)-\frac{1}{4}Y^2\right]
+\alpha^\prime-\beta^\prime
Q+\gamma^\prime\left[U(U+1)-\frac{1}{4}Q^2\right].
\end{equation}

In the case of flavor-spin $\SU(6)$-theory we have the following B\'{e}g-Singh mass formula \cite{BS,RF70}:
\begin{multline}\label{BS}
m^2=m^2_0+\mu_1C_2(3)+\mu_2\cdot 2J(J+1)+\mu_3Y+
\mu_4\left[2S(s+1)+\frac{1}{4}Y^2-C_2(4)\right]+\\
+\mu_5\left[-\frac{1}{2}Y^2+2T(T+1)\right]+
\mu_6\left[2N(N+1)-2S(S+1)\right],
\end{multline}
where $C_2(3)$ and $C_2(4)$ are Casimir operators of $\SU(6)$, $Y$ is a hypercharge, $J$ and $T$ are spin and isospin operators, $S$ and $N$ are so-called strange and non-strange spins. As in the case of $\SU(3)$-theory, a group action of $\SU(6)$ is analogous to Zeeman effect, that is, this action leads to a mass splitting of states within hypermultiplets of $\SU(6)$ (56-plet of baryons and 35-plet of mesons) via $\SU(6)/\SU(3)$- and $\SU(6)/\SU(4)$-reductions.

In the Gell-Mann--Okubo formulas (\ref{GMOQ})-(\ref{GMOL}) and B\'{e}g-Singh formula (\ref{BS}) $m_0$ depends on the chosen supermultiplet of $\SU(3)$ (hypermultiplet in the case of $\SU(6)$) and the concrete value of $m_0$ is not predicted by the $\SU(3)$- and $\SU(6)$-theories. Moreover, in Nature we see a wide variety of baryon octets (see, for example, Particle Data Group: pdg.lbl.gov), where mass distances between these octets are not explained by the $\SU(3)$- and $\SU(6)$-mass formulas. As a rule, all the predicted masses in $\SU(3)$- and $\SU(6)$-theories have a low accuracy (on an average 4\%--6\%).

It is well-known that in the standard model (SM) the number of basic parameters is 18, including the three gauge coupling constants \cite{Fri02}:
\[
m_e,\;m_u,\;m_d;\quad m_\mu,\;m_s,\;m_c;\quad m_\tau,\;m_b,\;m_t;
\]
\[
\theta_u,\;\theta_d,\;\theta,\;\delta;
\]
\[
M_w,\;M_h;
\]
\[
\alpha,\;\alpha_s,\alpha_W.
\]
Thirteen of these constants are directly related to the fermion masses, namely, lepton masses $(m_e,m_\mu,m_\tau)$, quark masses $(m_u,m_d,m_s,m_c,m_b,m_t)$ and mixing angles $(\theta_u,\theta_d,\theta,\delta)$. All these mass parameters have to be adjusted according to experimental measurements and cannot be predicted within the theory (SM). SM considers these mass parameters as ``fundamental constants''. For this reason the standard model must be regarded as a first step towards a more complete theoretical framework. As a consequence of this situation, we see in SM three types of so-called ``fundamental particles'' (quarks, leptons and gauge bosons).

On the other hand, in 1952, Numbu \cite{Num52} gave attention to an existence of empirical (``balmer-like'') relations in the mass spectrum of elementary particles:
\begin{equation}\label{Numbu}
m_N=\left(N/2\right)137\cdot m_e,
\end{equation}
where $N$ is a positive integer number, $m_e$ is the rest mass of electron. Further, in 1979, Barut \cite{Bar79} proposed a mass formula for the leptons:
\begin{equation}\label{Barut}
m(N)=m_e\left(1+\frac{3}{2}\alpha^{-1}\sum^{n=N}_{n=0}n^4\right),
\end{equation}
where $\alpha\approx 1/137$ is a fine structure constant. According to (\ref{Barut}), masses of electron, muon and $\tau$-lepton are defined at $N=0,1$ and 2, respectively. Later on, empirical relations of the form (\ref{Numbu}) were studied by many authors (see \cite{MG70,MG07,Pal07,SM08,MS08,Gro,Chiatti}). The Numbu formula (\ref{Numbu}) can be written also via the fine structure constant:
\begin{equation}\label{Alpha}
m=\frac{N}{2\alpha}m_e.
\end{equation}
This formula leads to a so-called $\alpha$-quantization of the elementary particle masses (see \cite{MG07,Gro}).

In 2003, Sidharth \cite{Sid1,Sid2} proposed the following empirical formula:
\begin{equation}\label{Sidharth}
\text{mass}=137\cdot m\left(n+\frac{1}{2}\right),
\end{equation}
where $m$ and $n$ are positive integer numbers. The Sidharth formula (\ref{Sidharth}) describes the all mass spectrum of elementary particles (known up to 2003) with an accuracy of $3\%$. Sidharth \cite{Sid2} attempted to relate the numbers $m$ and $n$ with the quantum numbers of harmonic oscillator. However, theoretical sense of these numbers, as the number $N$ in other ``balmer-like'' formulas ((\ref{Numbu})--(\ref{Alpha})), remains unclear.

In the present paper we study mass spectrum of ``elementary particles'', that is, a mass spectrum of actualized (localized) states of non-local quantum system (quantum domain). A spectrum of these states form a Hesenberg-like matter spectrum. According to Heisenberg \cite{Heisen1,Heisen}, in the ground of the all wide variety of elementary particles we have  \textit{substrate of energy}, the mathematical (group-theoretical) form of which (obtained via the fundamental symmetries) is a \textbf{\textit{matter spectrum}}. The each level (\textit{state}) of matter spectrum is defined by a representation of a group of fundamental symmetry. The each elementary particle presents itself a some energy level of this spectrum. An essential distinctive feature of such description is an absence of fundamental particles. Heisenberg claimed that a notion ``consists of'' does not work in particle physics. Applying this notion, we obtain that the each particle consists of the all known particles. For that reason among the all elementary particles we cannot to separate one as a fundamental particle \cite{Heisen}. In Heisenberg's approach we have \textit{fundamental symmetries} instead fundamental particles. In Heisenberg's opinion, all known symmetries in particle physics are divided on the two categories: \textit{fundamental (primary) symmetries} (such as the Lorentz group, discrete symmetries, conformal group) and \textit{dynamical (secondary) symmetries} (such as $\SU(3)$, $\SU(6)$ and so on). A quantum microobject is non-local and there exists as a superposition of state vectors of nonseparable Hilbert space outside the Minkowski space-time\footnote{It is interesting to note that in the well-known Penrose program \cite{Pen77,PM72} a twistor structure is understood as the underlying (more fundamental) structure with respect to Minkowski space-time. In other words, space-time continuum is not fundamental substance in the twistor approach, this is a fully derivative entity generated by the underlying twistor structure. In parallel with the twistor approach, decoherence theory \cite{JZKGKS} claims that in the background of reality we have a \textit{non-local quantum substrate} (quantum domain), and all visible world (classical domain$\equiv$space-time continuum) arises from quantum domain in the result of decoherence process \cite{Zur03,Zur03b}.}. A structure of the matter spectrum is organized within a Gelfand-Neumark-Segal construction \cite{GN43,Seg47}, where cyclic representations of the operator algebra (energy operator) are defined by fundamental symmetry (Lorentz group). Within the given realization of operator algebra\footnote{It is obvious that at the other realization of operator algebra (for example, via the conformal group as a fundamental symmetry) we obtain the other structure of matter spectrum.} we introduce a mass formula defining an energetic weight of the each level of matter spectrum. The obtained spectrum leads to a \textit{linearly} increasing mass spectrum of the states (``elementary particles''). Thus, all the states of matter spectrum are described as a \textbf{\textit{single quantum system}}\footnote{The linear character of mass spectrum (observed in Nature more than 60 years ago \cite{Num52}) is a direct consequence of the united quantum system.}. In the sections 3--6 we show that state masses of lepton and hadron sectors of matter spectrum are proportional to the rest mass of electron with an accuracy of $0,41\%$, that leads to a \textit{mass quantization}. This situation looks like a charge quantization (all charged states, observed in Nature, have a charge which is an integral multiple of the electron charge).

\section{Lorentz group and mass formula}
As is known, a universal covering of the proper orthochronous Lorentz group $\SO_0(1,3)$
(rotation group of the Minkowski space-time $\R^{1,3}$)
is the spinor group
\[\ar
\spin_+(1,3)\simeq\left\{\begin{pmatrix} \alpha & \beta \\ \gamma &
\delta
\end{pmatrix}\in\C_2:\;\;\det\begin{pmatrix}\alpha & \beta \\ \gamma & \delta
\end{pmatrix}=1\right\}=\SL(2,\C).
\]

Let $\fg\rightarrow T_{\fg}$ be an arbitrary linear
representation of the proper orthochronous Lorentz group
$\SO_0(1,3)$ and let $\sA_i(t)=T_{a_i(t)}$ be an infinitesimal
operator corresponding to the rotation $a_i(t)\in\SO_0(1,3)$.
Analogously, let $\sB_i(t)=T_{b_i(t)}$, where $b_i(t)\in\SO_0(1,3)$ is
the hyperbolic rotation. The elements $\sA_i$ and $\sB_i$ form a basis of the group algebra
$\mathfrak{sl}(2,\C)$ and satisfy the relations
\begin{equation}\label{Com1}
\left.\begin{array}{lll} \ld\sA_1,\sA_2\rd=\sA_3, &
\ld\sA_2,\sA_3\rd=\sA_1, &
\ld\sA_3,\sA_1\rd=\sA_2,\\[0.1cm]
\ld\sB_1,\sB_2\rd=-\sA_3, & \ld\sB_2,\sB_3\rd=-\sA_1, &
\ld\sB_3,\sB_1\rd=-\sA_2,\\[0.1cm]
\ld\sA_1,\sB_1\rd=0, & \ld\sA_2,\sB_2\rd=0, &
\ld\sA_3,\sB_3\rd=0,\\[0.1cm]
\ld\sA_1,\sB_2\rd=\sB_3, & \ld\sA_1,\sB_3\rd=-\sB_2, & \\[0.1cm]
\ld\sA_2,\sB_3\rd=\sB_1, & \ld\sA_2,\sB_1\rd=-\sB_3, & \\[0.1cm]
\ld\sA_3,\sB_1\rd=\sB_2, & \ld\sA_3,\sB_2\rd=-\sB_1. &
\end{array}\right\}
\end{equation}
Defining the operators
\begin{gather}
\sX_l=\frac{1}{2}i(\sA_l+i\sB_l),\quad\sY_l=\frac{1}{2}i(\sA_l-i\sB_l),
\label{SL25}\\ 
(l=1,2,3),\nonumber
\end{gather}
we come to a \textit{complex envelope} of the group algebra $\mathfrak{sl}(2,\C)$.
Using the relations (\ref{Com1}), we find
\begin{equation}\label{Com2}
\ld\sX_k,\sX_l\rd=i\varepsilon_{klm}\sX_m,\quad
\ld\sY_l,\sY_m\rd=i\varepsilon_{lmn}\sY_n,\quad \ld\sX_l,\sY_m\rd=0.
\end{equation}
From the relations (\ref{Com2}) it follows that each of the sets of
infinitesimal operators $\sX$ and $\sY$ generates the group $\SU(2)$
and these two groups commute with each other. Thus, from the
relations (\ref{Com2}) it follows that the group algebra
$\mathfrak{sl}(2,\C)$ (within the complex envelope) is algebraically isomorphic to the direct sum $\mathfrak{su}(2)\oplus\mathfrak{su}(2)$\footnote{In a sense, it allows one to represent the group
$\SL(2,\C)$ by a product $\SU(2)\otimes\SU(2)$ as it done by Ryder in his
textbook \cite{Ryd85}. Moreover, in the works \cite{AE93,Dvo96} the
Lorentz group is represented by a product $\SU_R(2)\otimes\SU_L(2)$,
where the spinors $\psi(p^\mu)=\begin{pmatrix}\phi_R(p^\mu)\\
\phi_L(p^\mu)\end{pmatrix}$ ($\phi_R(p^\mu)$ and $\phi_L(p^\mu)$ are
the right- and left-handed spinors) are transformed within
$(j,0)\oplus(0,j)$ representation space, in our case $j=l=\dot{l}$.
However, the isomorphism $\SL(2,C)\simeq\SU(2)\otimes\SU(2)$ is not correct from group-theoretical viewpoint. Indeed, the groups $\SO_0(1,3)$ and $\SO(4)$ are real forms of the complex 6-dimensional Lie group $\SO(4,\C)$ with complex Lie algebra $D_2=A_1+A_1$. Real Lie algebras are compact iff the Killing form is negative definite \cite{Knapp}. That is the case for Lie algebra of $\SO(4)$, not for $\SO_0(1,3)$.}.

Further, introducing operators of the form (`rising' and `lowering' operators of the group $\SL(2,\C)$)
\begin{equation}\label{SL26}
\left.\begin{array}{cc}
\sX_+=\sX_1+i\sX_2, & \sX_-=\sX_1-i\sX_2,\\[0.1cm]
\sY_+=\sY_1+i\sY_2, & \sY_-=\sY_1-i\sY_2,
\end{array}\right\}
\end{equation}
we see that
\[
\ld\sX_3,\sX_+\rd=\sX_+,\quad\ld\sX_3,\sX_-\rd=-\sX_-,\quad\ld\sX_+,\sX_-\rd=2\sX_3,
\]
\[
\ld\sY_3,\sY_+\rd=\sY_+,\quad\ld\sY_3,\sY_-\rd=-\sY_-,\quad\ld\sY_+,\sY_-\rd=2\sY_3.
\]


\unitlength=1.5mm
\begin{center}
\begin{picture}(105,70)
\put(50,0){$\overset{(0,0)}{\bullet}$}\put(47,5.5){\line(1,0){10}}\put(52.25,2.75){\line(0,1){7.25}}
\put(55,5){$\overset{(\frac{1}{2},0)}{\bullet}$}
\put(45,5){$\overset{(0,\frac{1}{2})}{\bullet}$}
\put(40,10){$\overset{(0,1)}{\bullet}$}\put(42,10.5){\line(1,0){10}}\put(47.25,7.75){\line(0,1){7.25}}
\put(50,10){$\overset{(\frac{1}{2},\frac{1}{2})}{\bullet}$}
\put(52,10.5){\line(1,0){10}}\put(57.25,7.75){\line(0,1){7.25}}
\put(60,10){$\overset{(1,0)}{\bullet}$}
\put(35,15){$\overset{(0,\frac{3}{2})}{\bullet}$}\put(37,15.5){\line(1,0){10}}\put(42.25,12.75){\line(0,1){7.25}}
\put(45,15){$\overset{(\frac{1}{2},1)}{\bullet}$}\put(47,15.5){\line(1,0){10}}\put(52.25,12.75){\line(0,1){7.25}}
\put(55,15){$\overset{(1,\frac{1}{2})}{\bullet}$}\put(57,15.5){\line(1,0){10}}\put(62.25,12.75){\line(0,1){7.25}}
\put(65,15){$\overset{(\frac{3}{2},0)}{\bullet}$}
\put(30,20){$\overset{(0,2)}{\bullet}$}\put(32,20.5){\line(1,0){10}}\put(37.25,17.75){\line(0,1){7.25}}
\put(40,20){$\overset{(\frac{1}{2},\frac{3}{2})}{\bullet}$}
\put(42,20.5){\line(1,0){10}}\put(47.25,17.75){\line(0,1){7.25}}
\put(50,20){$\overset{(1,1)}{\bullet}$}\put(52,20.5){\line(1,0){10}}\put(57.25,17.75){\line(0,1){7.25}}
\put(60,20){$\overset{(\frac{3}{2},\frac{1}{2})}{\bullet}$}
\put(62,20.5){\line(1,0){10}}\put(67.25,17.75){\line(0,1){7.25}}
\put(70,20){$\overset{(2,0)}{\bullet}$}
\put(25,25){$\overset{(0,\frac{5}{2})}{\bullet}$}\put(27,25.5){\line(1,0){10}}\put(32.25,22.75){\line(0,1){7.25}}
\put(35,25){$\overset{(\frac{1}{2},2)}{\bullet}$}\put(37,25.5){\line(1,0){10}}\put(42.25,22.75){\line(0,1){7.25}}
\put(45,25){$\overset{(1,\frac{3}{2})}{\bullet}$}\put(47,25.5){\line(1,0){10}}\put(52.25,22.75){\line(0,1){7.25}}
\put(55,25){$\overset{(\frac{3}{2},1)}{\bullet}$}\put(57,25.5){\line(1,0){10}}\put(62.25,22.75){\line(0,1){7.25}}
\put(65,25){$\overset{(2,\frac{1}{2})}{\bullet}$}\put(67,25.5){\line(1,0){10}}\put(72.25,22.75){\line(0,1){7.25}}
\put(75,25){$\overset{(\frac{5}{2},0)}{\bullet}$}
\put(20,30){$\overset{(0,3)}{\bullet}$}\put(22,30.5){\line(1,0){10}}\put(27.25,27.75){\line(0,1){7.25}}
\put(30,30){$\overset{(\frac{1}{2},\frac{5}{2})}{\bullet}$}
\put(32,30.5){\line(1,0){10}}\put(37.25,27.75){\line(0,1){7.25}}
\put(40,30){$\overset{(1,2)}{\bullet}$}\put(42,30.5){\line(1,0){10}}\put(47.25,27.75){\line(0,1){7.25}}
\put(50,30){$\overset{(\frac{3}{2},\frac{3}{2})}{\bullet}$}
\put(52,30.5){\line(1,0){10}}\put(57.25,27.75){\line(0,1){7.25}}
\put(60,30){$\overset{(2,1)}{\bullet}$}\put(62,30.5){\line(1,0){10}}\put(67.25,27.75){\line(0,1){7.25}}
\put(70,30){$\overset{(\frac{5}{2},\frac{5}{2})}{\bullet}$}
\put(72,30.5){\line(1,0){10}}\put(77.25,27.75){\line(0,1){7.25}}
\put(80,30){$\overset{(3,0)}{\bullet}$}
\put(15,35){$\overset{(0,\frac{7}{2})}{\bullet}$}\put(17,35.5){\line(1,0){10}}\put(22.25,32.75){\line(0,1){7.25}}
\put(25,35){$\overset{(\frac{1}{2},3)}{\bullet}$}\put(27,35.5){\line(1,0){10}}\put(32.25,32.75){\line(0,1){7.25}}
\put(35,35){$\overset{(1,\frac{5}{2})}{\bullet}$}\put(37,35.5){\line(1,0){10}}\put(42.25,32.75){\line(0,1){7.25}}
\put(45,35){$\overset{(\frac{3}{2},2)}{\bullet}$}\put(47,35.5){\line(1,0){10}}\put(52.25,32.75){\line(0,1){7.25}}
\put(55,35){$\overset{(2,\frac{3}{2})}{\bullet}$}\put(57,35.5){\line(1,0){10}}\put(62.25,32.75){\line(0,1){7.25}}
\put(65,35){$\overset{(\frac{5}{2},1)}{\bullet}$}\put(67,35.5){\line(1,0){10}}\put(72.25,32.75){\line(0,1){7.25}}
\put(75,35){$\overset{(3,\frac{1}{2})}{\bullet}$}\put(77,35.5){\line(1,0){10}}\put(82.25,32.75){\line(0,1){7.25}}
\put(85,35){$\overset{(\frac{7}{2},0)}{\bullet}$}
\put(10,40){$\overset{(0,4)}{\bullet}$}\put(12,40.5){\line(1,0){10}}
\put(5,45){$\overset{(0,\frac{9}{2})}{\bullet}$}
\put(15,45){$\overset{(\frac{1}{2},4)}{\bullet}$}
\put(25,45){$\overset{(1,\frac{7}{2})}{\bullet}$}
\put(35,45){$\overset{(\frac{3}{2},3)}{\bullet}$}
\put(45,45){$\overset{(2,\frac{5}{2})}{\bullet}$}
\put(55,45){$\overset{(\frac{5}{2},2)}{\bullet}$}
\put(65,45){$\overset{(3,\frac{3}{2})}{\bullet}$}
\put(75,45){$\overset{(\frac{7}{2},1)}{\bullet}$}
\put(85,45){$\overset{(4,\frac{1}{2})}{\bullet}$}
\put(0,50){$\overset{(0,5)}{\bullet}$}
\put(10,50){$\overset{(\frac{1}{2},\frac{9}{2})}{\bullet}$}
\put(20,50){$\overset{(1,4)}{\bullet}$}
\put(30,50){$\overset{(\frac{3}{2},\frac{7}{2})}{\bullet}$}
\put(40,50){$\overset{(2,3)}{\bullet}$}
\put(50,50){$\overset{(\frac{5}{2},\frac{5}{2})}{\bullet}$}
\put(60,50){$\overset{(3,2)}{\bullet}$}
\put(70,50){$\overset{(\frac{7}{2},\frac{3}{2})}{\bullet}$}
\put(80,50){$\overset{(4,1)}{\bullet}$}
\put(90,50){$\overset{(\frac{9}{2},\frac{1}{2})}{\bullet}$}
\put(-5,55){$\overset{(0,\frac{11}{2})}{\bullet}$}
\put(5,55){$\overset{(\frac{1}{2},5)}{\bullet}$}
\put(15,55){$\overset{(1,\frac{9}{2})}{\bullet}$}
\put(25,55){$\overset{(\frac{3}{2},4)}{\bullet}$}
\put(35,55){$\overset{(2,\frac{7}{2})}{\bullet}$}
\put(45,55){$\overset{(\frac{5}{2},3)}{\bullet}$}
\put(55,55){$\overset{(3,\frac{5}{2})}{\bullet}$}
\put(65,55){$\overset{(\frac{7}{2},2)}{\bullet}$}
\put(75,55){$\overset{(4,\frac{3}{2})}{\bullet}$}
\put(85,55){$\overset{(\frac{9}{2},1)}{\bullet}$}
\put(95,55){$\overset{(5,\frac{1}{2})}{\bullet}$}
\put(20,40){$\overset{(\frac{1}{2},\frac{7}{2})}{\bullet}$}\put(22,40.5){\line(1,0){10}}
\put(7.25,47.75){\line(0,1){7.25}}
\put(17.25,47.75){\line(0,1){7.25}}
\put(17.25,37.75){\line(0,1){7.25}}
\put(27.25,47.75){\line(0,1){7.25}}
\put(27.25,37.75){\line(0,1){7.25}}
\put(37.25,47.75){\line(0,1){7.25}}
\put(37.25,37.75){\line(0,1){7.25}}
\put(47.25,47.75){\line(0,1){7.25}}
\put(47.25,37.75){\line(0,1){7.25}}
\put(57.25,47.75){\line(0,1){7.25}}
\put(57.25,37.75){\line(0,1){7.25}}
\put(67.25,47.75){\line(0,1){7.25}}
\put(67.25,37.75){\line(0,1){7.25}}
\put(77.25,47.75){\line(0,1){7.25}}
\put(77.25,37.75){\line(0,1){7.25}}
\put(87.25,47.75){\line(0,1){7.25}}
\put(87.25,37.75){\line(0,1){7.25}}
\put(97.25,47.75){\line(0,1){7.25}}
\put(12.25,42.75){\line(0,1){7.25}}
\put(22.25,42.75){\line(0,1){7.25}}
\put(32.25,42.75){\line(0,1){7.25}}
\put(42.25,42.75){\line(0,1){7.25}}
\put(52.25,42.75){\line(0,1){7.25}}
\put(62.25,42.75){\line(0,1){7.25}}
\put(72.25,42.75){\line(0,1){7.25}}
\put(82.25,42.75){\line(0,1){7.25}}
\put(92.25,42.75){\line(0,1){7.25}}
\put(2,50.5){\line(1,0){10}}
\put(12,50.5){\line(1,0){10}}
\put(22,50.5){\line(1,0){10}}
\put(32,50.5){\line(1,0){10}}
\put(42,50.5){\line(1,0){10}}
\put(52,50.5){\line(1,0){10}}
\put(62,50.5){\line(1,0){10}}
\put(72,50.5){\line(1,0){10}}
\put(82,50.5){\line(1,0){10}}
\put(92,50.5){\line(1,0){10}}
\put(7,45.5){\line(1,0){10}}
\put(17,45.5){\line(1,0){10}}
\put(27,45.5){\line(1,0){10}}
\put(37,45.5){\line(1,0){10}}
\put(47,45.5){\line(1,0){10}}
\put(57,45.5){\line(1,0){10}}
\put(67,45.5){\line(1,0){10}}
\put(77,45.5){\line(1,0){10}}
\put(87,45.5){\line(1,0){10}}
\put(-2.75,55.5){\line(1,0){10}}
\put(7.25,55.5){\line(1,0){10}}
\put(17.25,55.5){\line(1,0){10}}
\put(27.25,55.5){\line(1,0){10}}
\put(37.25,55.5){\line(1,0){10}}
\put(47.25,55.5){\line(1,0){10}}
\put(57.25,55.5){\line(1,0){10}}
\put(67.25,55.5){\line(1,0){10}}
\put(77.25,55.5){\line(1,0){10}}
\put(87.25,55.5){\line(1,0){10}}
\put(97.25,55.5){\line(1,0){10}}
\put(30,40){$\overset{(1,3)}{\bullet}$}\put(32,40.5){\line(1,0){10}}
\put(40,40){$\overset{(\frac{3}{2},\frac{5}{2})}{\bullet}$}\put(42,40.5){\line(1,0){10}}
\put(50,40){$\overset{(2,2)}{\bullet}$}\put(52,40.5){\line(1,0){10}}
\put(60,40){$\overset{(\frac{5}{2},\frac{3}{2})}{\bullet}$}\put(62,40.5){\line(1,0){10}}
\put(70,40){$\overset{(3,1)}{\bullet}$}\put(72,40.5){\line(1,0){10}}
\put(80,40){$\overset{(\frac{7}{2},\frac{1}{2})}{\bullet}$}\put(82,40.5){\line(1,0){10}}
\put(90,40){$\overset{(4,0)}{\bullet}$}
\put(95,45){$\overset{(\frac{9}{2},0)}{\bullet}$}
\put(100,50){$\overset{(5,0)}{\bullet}$}
\put(105,55){$\overset{(\frac{11}{2},0)}{\bullet}$}
\put(-2.5,60){$\vdots$}
\put(6.5,60){$\vdots$}
\put(16.5,60){$\vdots$}
\put(26.5,60){$\vdots$}
\put(36.5,60){$\vdots$}
\put(46.5,60){$\vdots$}
\put(56.5,60){$\vdots$}
\put(66.5,60){$\vdots$}
\put(76.5,60){$\vdots$}
\put(86.5,60){$\vdots$}
\put(96.5,60){$\vdots$}
\put(106.5,60){$\vdots$}
\put(-4,0.5){\line(1,0){55}}\put(50,0.5){\vector(1,0){57}}
\put(1.5,49){$\vdots$}
\put(1.5,46){$\vdots$}
\put(1.5,43){$\vdots$}
\put(1.5,41){$\vdots$}
\put(1.5,38){$\vdots$}
\put(1.5,35){$\vdots$}
\put(1.5,32){$\vdots$}
\put(1.5,29){$\vdots$}
\put(1.5,26){$\vdots$}
\put(1.5,23){$\vdots$}
\put(1.5,20){$\vdots$}
\put(1.5,17){$\vdots$}
\put(1.5,14){$\vdots$}
\put(1.5,11){$\vdots$}
\put(1.5,9){$\vdots$}
\put(1.5,6){$\vdots$}
\put(1.5,3){$\vdots$}
\put(1.5,1.5){$\cdot$}
\put(1.5,0){$\cdot$}
\put(-0.5,-2){$-5$}
\put(6.5,43){$\vdots$}
\put(6.5,41){$\vdots$}
\put(6.5,38){$\vdots$}
\put(6.5,35){$\vdots$}
\put(6.5,32){$\vdots$}
\put(6.5,29){$\vdots$}
\put(6.5,26){$\vdots$}
\put(6.5,23){$\vdots$}
\put(6.5,20){$\vdots$}
\put(6.5,17){$\vdots$}
\put(6.5,14){$\vdots$}
\put(6.5,11){$\vdots$}
\put(6.5,9){$\vdots$}
\put(6.5,6){$\vdots$}
\put(6.5,3){$\vdots$}
\put(6.5,1.5){$\cdot$}
\put(6.5,0){$\cdot$}
\put(4.5,-2){$-\frac{9}{2}$}
\put(11.5,38){$\vdots$}
\put(11.5,35){$\vdots$}
\put(11.5,32){$\vdots$}
\put(11.5,29){$\vdots$}
\put(11.5,26){$\vdots$}
\put(11.5,23){$\vdots$}
\put(11.5,20){$\vdots$}
\put(11.5,17){$\vdots$}
\put(11.5,14){$\vdots$}
\put(11.5,11){$\vdots$}
\put(11.5,9){$\vdots$}
\put(11.5,6){$\vdots$}
\put(11.5,3){$\vdots$}
\put(11.5,1.5){$\cdot$}
\put(11.5,0){$\cdot$}
\put(9.5,-2){$-4$}
\put(16.5,32){$\vdots$}
\put(16.5,29){$\vdots$}
\put(16.5,26){$\vdots$}
\put(16.5,23){$\vdots$}
\put(16.5,20){$\vdots$}
\put(16.5,17){$\vdots$}
\put(16.5,14){$\vdots$}
\put(16.5,11){$\vdots$}
\put(16.5,9){$\vdots$}
\put(16.5,6){$\vdots$}
\put(16.5,3){$\vdots$}
\put(16.5,1.5){$\cdot$}
\put(16.5,0){$\cdot$}
\put(14.5,-2){$-\frac{7}{2}$}
\put(21.5,27){$\vdots$}
\put(21.5,24){$\vdots$}
\put(21.5,21){$\vdots$}
\put(21.5,18){$\vdots$}
\put(21.5,15){$\vdots$}
\put(21.5,13){$\vdots$}
\put(21.5,9){$\vdots$}
\put(21.5,6){$\vdots$}
\put(21.5,3){$\vdots$}
\put(21.5,1.5){$\cdot$}
\put(21.5,0){$\cdot$}
\put(19.5,-2){$-3$}
\put(26.5,22){$\vdots$}
\put(26.5,19){$\vdots$}
\put(26.5,16){$\vdots$}
\put(26.5,13){$\vdots$}
\put(26.5,10){$\vdots$}
\put(26.5,7){$\vdots$}
\put(26.5,4){$\vdots$}
\put(26.5,1){$\vdots$}
\put(24.5,-2){$-\frac{5}{2}$}
\put(31.5,17){$\vdots$}
\put(31.5,14){$\vdots$}
\put(31.5,11){$\vdots$}
\put(31.5,8){$\vdots$}
\put(31.5,5){$\vdots$}
\put(31.5,2){$\vdots$}
\put(31.5,0.5){$\cdot$}
\put(29.5,-2){$-2$}
\put(36.5,12){$\vdots$}
\put(36.5,9){$\vdots$}
\put(36.5,6){$\vdots$}
\put(36.5,3){$\vdots$}
\put(36.5,1.5){$\cdot$}
\put(36.5,0){$\cdot$}
\put(34.5,-2){$-\frac{3}{2}$}
\put(41.5,7){$\vdots$}
\put(41.5,4){$\vdots$}
\put(41.5,1){$\vdots$}
\put(39.5,-2){$-1$}
\put(46.5,2){$\vdots$}
\put(46.5,0.5){$\cdot$}
\put(44.5,-2){$-\frac{1}{2}$}
\put(51.5,-2){$0$}
\put(56.5,2){$\vdots$}
\put(56.5,0.5){$\cdot$}
\put(56.5,-2){$\frac{1}{2}$}
\put(61.5,7){$\vdots$}
\put(61.5,4){$\vdots$}
\put(61.5,1){$\vdots$}
\put(61.5,-2){$1$}
\put(66.5,12){$\vdots$}
\put(66.5,9){$\vdots$}
\put(66.5,6){$\vdots$}
\put(66.5,3){$\vdots$}
\put(66.5,1.5){$\cdot$}
\put(66.5,0){$\cdot$}
\put(66.5,-2){$\frac{3}{2}$}
\put(71.5,17){$\vdots$}
\put(71.5,14){$\vdots$}
\put(71.5,11){$\vdots$}
\put(71.5,8){$\vdots$}
\put(71.5,5){$\vdots$}
\put(71.5,2){$\vdots$}
\put(71.5,0.5){$\cdot$}
\put(71.5,-2){$2$}
\put(76.5,22){$\vdots$}
\put(76.5,19){$\vdots$}
\put(76.5,16){$\vdots$}
\put(76.5,13){$\vdots$}
\put(76.5,10){$\vdots$}
\put(76.5,7){$\vdots$}
\put(76.5,4){$\vdots$}
\put(76.5,1){$\vdots$}
\put(76.5,-2){$\frac{5}{2}$}
\put(81.5,27){$\vdots$}
\put(81.5,24){$\vdots$}
\put(81.5,21){$\vdots$}
\put(81.5,18){$\vdots$}
\put(81.5,15){$\vdots$}
\put(81.5,13){$\vdots$}
\put(81.5,9){$\vdots$}
\put(81.5,6){$\vdots$}
\put(81.5,3){$\vdots$}
\put(81.5,1.5){$\cdot$}
\put(81.5,0){$\cdot$}
\put(81.5,-2){$3$}
\put(86.5,32){$\vdots$}
\put(86.5,29){$\vdots$}
\put(86.5,26){$\vdots$}
\put(86.5,23){$\vdots$}
\put(86.5,20){$\vdots$}
\put(86.5,17){$\vdots$}
\put(86.5,14){$\vdots$}
\put(86.5,11){$\vdots$}
\put(86.5,9){$\vdots$}
\put(86.5,6){$\vdots$}
\put(86.5,3){$\vdots$}
\put(86.5,1.5){$\cdot$}
\put(86.5,0){$\cdot$}
\put(86.5,-2){$\frac{7}{2}$}
\put(91.5,38){$\vdots$}
\put(91.5,35){$\vdots$}
\put(91.5,32){$\vdots$}
\put(91.5,29){$\vdots$}
\put(91.5,26){$\vdots$}
\put(91.5,23){$\vdots$}
\put(91.5,20){$\vdots$}
\put(91.5,17){$\vdots$}
\put(91.5,14){$\vdots$}
\put(91.5,11){$\vdots$}
\put(91.5,9){$\vdots$}
\put(91.5,6){$\vdots$}
\put(91.5,3){$\vdots$}
\put(91.5,1.5){$\cdot$}
\put(91.5,0){$\cdot$}
\put(91.5,-2){$4$}
\put(96.5,44){$\vdots$}
\put(96.5,41){$\vdots$}
\put(96.5,38){$\vdots$}
\put(96.5,35){$\vdots$}
\put(96.5,32){$\vdots$}
\put(96.5,29){$\vdots$}
\put(96.5,26){$\vdots$}
\put(96.5,23){$\vdots$}
\put(96.5,20){$\vdots$}
\put(96.5,17){$\vdots$}
\put(96.5,14){$\vdots$}
\put(96.5,11){$\vdots$}
\put(96.5,9){$\vdots$}
\put(96.5,6){$\vdots$}
\put(96.5,3){$\vdots$}
\put(96.5,1.5){$\cdot$}
\put(96.5,0){$\cdot$}
\put(96.5,-2){$\frac{9}{2}$}
\put(101.5,50){$\vdots$}
\put(101.5,47){$\vdots$}
\put(101.5,44){$\vdots$}
\put(101.5,41){$\vdots$}
\put(101.5,38){$\vdots$}
\put(101.5,35){$\vdots$}
\put(101.5,32){$\vdots$}
\put(101.5,29){$\vdots$}
\put(101.5,26){$\vdots$}
\put(101.5,23){$\vdots$}
\put(101.5,20){$\vdots$}
\put(101.5,17){$\vdots$}
\put(101.5,14){$\vdots$}
\put(101.5,11){$\vdots$}
\put(101.5,9){$\vdots$}
\put(101.5,6){$\vdots$}
\put(101.5,3){$\vdots$}
\put(101.5,1.5){$\cdot$}
\put(101.5,0){$\cdot$}
\put(101.5,-2){$5$}
\put(53.8,1.7){$\cdot$}\put(54.3,2.2){$\cdot$}\put(54.8,2.7){$\cdot$}\put(55.3,3.3){$\cdot$}\put(55.8,3.8){$\cdot$}
\put(56.3,4.3){$\cdot$}
\put(58.8,6.8){$\cdot$}\put(59.3,7.3){$\cdot$}\put(59.8,7.8){$\cdot$}\put(60.3,8.3){$\cdot$}\put(60.8,8.8){$\cdot$}
\put(61.3,9.3){$\cdot$}
\put(63.8,11.8){$\cdot$}\put(64.3,12.3){$\cdot$}\put(64.8,12.8){$\cdot$}\put(65.3,13.3){$\cdot$}\put(65.8,13.8){$\cdot$}
\put(66.3,14.3){$\cdot$}
\put(68.8,16.8){$\cdot$}\put(69.3,17.3){$\cdot$}\put(69.8,17.8){$\cdot$}\put(70.3,18.3){$\cdot$}\put(70.8,18.8){$\cdot$}
\put(71.3,19.3){$\cdot$}
\put(33.8,21.8){$\cdot$}\put(34.3,22.3){$\cdot$}\put(34.8,22.8){$\cdot$}\put(35.3,23.3){$\cdot$}\put(35.8,23.8){$\cdot$}
\put(36.3,24.3){$\cdot$}
\put(38.8,26.8){$\cdot$}\put(39.3,27.3){$\cdot$}\put(39.8,27.8){$\cdot$}\put(40.3,28.3){$\cdot$}\put(40.8,28.8){$\cdot$}
\put(41.3,29.3){$\cdot$}
\put(43.8,31.8){$\cdot$}\put(44.3,32.3){$\cdot$}\put(44.8,32.8){$\cdot$}\put(45.3,33.3){$\cdot$}\put(45.8,33.8){$\cdot$}
\put(46.3,34.3){$\cdot$}
\put(48.8,36.8){$\cdot$}\put(49.3,37.3){$\cdot$}\put(49.8,37.8){$\cdot$}\put(50.3,38.3){$\cdot$}\put(50.8,38.8){$\cdot$}
\put(51.3,39.3){$\cdot$}
\put(47.3,4.4){$\cdot$}\put(47.8,3.9){$\cdot$}\put(48.3,3.4){$\cdot$}\put(48.8,2.9){$\cdot$}\put(49.3,2.4){$\cdot$}
\put(49.8,1.9){$\cdot$}
\put(42.3,9.4){$\cdot$}\put(42.8,8.9){$\cdot$}\put(43.3,8.4){$\cdot$}\put(43.8,7.9){$\cdot$}\put(44.3,7.4){$\cdot$}
\put(44.8,6.9){$\cdot$}
\put(37.3,14.4){$\cdot$}\put(37.8,13.9){$\cdot$}\put(38.3,13.4){$\cdot$}\put(38.8,12.9){$\cdot$}\put(39.3,12.4){$\cdot$}
\put(39.8,11.9){$\cdot$}
\put(32.3,19.4){$\cdot$}\put(32.8,18.9){$\cdot$}\put(33.3,18.4){$\cdot$}\put(33.8,17.9){$\cdot$}\put(34.3,17.4){$\cdot$}
\put(34.8,16.9){$\cdot$}
\put(67.3,24.4){$\cdot$}\put(67.8,23.9){$\cdot$}\put(68.3,23.4){$\cdot$}\put(68.8,22.9){$\cdot$}\put(69.3,22.4){$\cdot$}
\put(69.8,21.9){$\cdot$}
\put(62.3,29.4){$\cdot$}\put(62.8,28.9){$\cdot$}\put(63.3,28.4){$\cdot$}\put(63.8,27.9){$\cdot$}\put(64.3,27.4){$\cdot$}
\put(64.8,26.9){$\cdot$}
\put(57.3,34.4){$\cdot$}\put(57.8,33.9){$\cdot$}\put(58.3,33.4){$\cdot$}\put(58.8,32.9){$\cdot$}\put(59.3,32.4){$\cdot$}
\put(59.8,31.9){$\cdot$}
\put(52.3,39.4){$\cdot$}\put(52.8,38.9){$\cdot$}\put(53.3,38.4){$\cdot$}\put(53.8,37.9){$\cdot$}\put(54.3,37.4){$\cdot$}
\put(54.8,36.9){$\cdot$}
\put(73.8,21.8){$\cdot$}\put(74.3,22.3){$\cdot$}\put(74.8,22.8){$\cdot$}\put(75.3,23.3){$\cdot$}\put(75.8,23.8){$\cdot$}
\put(76.3,24.3){$\cdot$}
\put(78.8,26.8){$\cdot$}\put(79.3,27.3){$\cdot$}\put(79.8,27.8){$\cdot$}\put(80.3,28.3){$\cdot$}\put(80.8,28.8){$\cdot$}
\put(81.3,29.3){$\cdot$}
\put(83.8,31.8){$\cdot$}\put(84.3,32.3){$\cdot$}\put(84.8,32.8){$\cdot$}\put(85.3,33.3){$\cdot$}\put(85.8,33.8){$\cdot$}
\put(86.3,34.3){$\cdot$}
\put(88.8,36.8){$\cdot$}\put(89.3,37.3){$\cdot$}\put(89.8,37.8){$\cdot$}\put(90.3,38.3){$\cdot$}\put(90.8,38.8){$\cdot$}
\put(91.3,39.3){$\cdot$}
\put(93.8,41.8){$\cdot$}\put(94.3,42.3){$\cdot$}\put(94.8,42.8){$\cdot$}\put(95.3,43.3){$\cdot$}\put(95.8,43.8){$\cdot$}
\put(96.3,44.3){$\cdot$}
\put(98.8,46.8){$\cdot$}\put(99.3,47.3){$\cdot$}\put(99.8,47.8){$\cdot$}\put(100.3,48.3){$\cdot$}\put(100.8,48.8){$\cdot$}
\put(101.3,49.3){$\cdot$}
\put(103.8,51.8){$\cdot$}\put(104.3,52.3){$\cdot$}\put(104.8,52.8){$\cdot$}\put(105.3,53.3){$\cdot$}\put(105.8,53.8){$\cdot$}
\put(106.3,54.3){$\cdot$}
\put(27.3,24.4){$\cdot$}\put(27.8,23.9){$\cdot$}\put(28.3,23.4){$\cdot$}\put(28.8,22.9){$\cdot$}\put(29.3,22.4){$\cdot$}
\put(29.8,21.9){$\cdot$}
\put(22.3,29.4){$\cdot$}\put(22.8,28.9){$\cdot$}\put(23.3,28.4){$\cdot$}\put(23.8,27.9){$\cdot$}\put(24.3,27.4){$\cdot$}
\put(24.8,26.9){$\cdot$}
\put(17.3,34.4){$\cdot$}\put(17.8,33.9){$\cdot$}\put(18.3,33.4){$\cdot$}\put(18.8,32.9){$\cdot$}\put(19.3,32.4){$\cdot$}
\put(19.8,31.9){$\cdot$}
\put(12.3,39.4){$\cdot$}\put(12.8,38.9){$\cdot$}\put(13.3,38.4){$\cdot$}\put(13.8,37.9){$\cdot$}\put(14.3,37.4){$\cdot$}
\put(14.8,36.9){$\cdot$}
\put(7.3,44.4){$\cdot$}\put(7.8,43.9){$\cdot$}\put(8.3,43.4){$\cdot$}\put(8.8,42.9){$\cdot$}\put(9.3,42.4){$\cdot$}
\put(9.8,41.9){$\cdot$}
\put(2.3,49.4){$\cdot$}\put(2.8,48.9){$\cdot$}\put(3.3,48.4){$\cdot$}\put(3.8,47.9){$\cdot$}\put(4.3,47.4){$\cdot$}
\put(4.8,46.9){$\cdot$}
\put(-2.7,54.4){$\cdot$}\put(-2.2,53.9){$\cdot$}\put(-1.7,53.4){$\cdot$}\put(-1.2,52.9){$\cdot$}\put(-0.7,52.4){$\cdot$}
\put(-0.2,51.9){$\cdot$}
\put(13.8,41.8){$\cdot$}\put(14.3,42.3){$\cdot$}\put(14.8,42.8){$\cdot$}\put(15.3,43.3){$\cdot$}\put(15.8,43.8){$\cdot$}
\put(16.3,44.3){$\cdot$}
\put(18.8,46.8){$\cdot$}\put(19.3,47.3){$\cdot$}\put(19.8,47.8){$\cdot$}\put(20.3,48.3){$\cdot$}\put(20.8,48.8){$\cdot$}
\put(21.3,49.3){$\cdot$}
\put(23.8,51.8){$\cdot$}\put(24.3,52.3){$\cdot$}\put(24.8,52.8){$\cdot$}\put(25.3,53.3){$\cdot$}\put(25.8,53.8){$\cdot$}
\put(26.3,54.3){$\cdot$}
\put(53.8,41.8){$\cdot$}\put(54.3,42.3){$\cdot$}\put(54.8,42.8){$\cdot$}\put(55.3,43.3){$\cdot$}\put(55.8,43.8){$\cdot$}
\put(56.3,44.3){$\cdot$}
\put(58.8,46.8){$\cdot$}\put(59.3,47.3){$\cdot$}\put(59.8,47.8){$\cdot$}\put(60.3,48.3){$\cdot$}\put(60.8,48.8){$\cdot$}
\put(61.3,49.3){$\cdot$}
\put(63.8,51.8){$\cdot$}\put(64.3,52.3){$\cdot$}\put(64.8,52.8){$\cdot$}\put(65.3,53.3){$\cdot$}\put(65.8,53.8){$\cdot$}
\put(66.3,54.3){$\cdot$}
\put(47.3,44.4){$\cdot$}\put(47.8,43.9){$\cdot$}\put(48.3,43.4){$\cdot$}\put(48.8,42.9){$\cdot$}\put(49.3,42.4){$\cdot$}
\put(49.8,41.9){$\cdot$}
\put(42.3,49.4){$\cdot$}\put(42.8,48.9){$\cdot$}\put(43.3,48.4){$\cdot$}\put(43.8,47.9){$\cdot$}\put(44.3,47.4){$\cdot$}
\put(44.8,46.9){$\cdot$}
\put(37.3,54.4){$\cdot$}\put(37.8,53.9){$\cdot$}\put(38.3,53.4){$\cdot$}\put(38.8,52.9){$\cdot$}\put(39.3,52.4){$\cdot$}
\put(39.8,51.9){$\cdot$}
\put(87.3,44.4){$\cdot$}\put(87.8,43.9){$\cdot$}\put(88.3,43.4){$\cdot$}\put(88.8,42.9){$\cdot$}\put(89.3,42.4){$\cdot$}
\put(89.8,41.9){$\cdot$}
\put(82.3,49.4){$\cdot$}\put(82.8,48.9){$\cdot$}\put(83.3,48.4){$\cdot$}\put(83.8,47.9){$\cdot$}\put(84.3,47.4){$\cdot$}
\put(84.8,46.9){$\cdot$}
\put(77.3,54.4){$\cdot$}\put(77.8,53.9){$\cdot$}\put(78.3,53.4){$\cdot$}\put(78.8,52.9){$\cdot$}\put(79.3,52.4){$\cdot$}
\put(79.8,51.9){$\cdot$}
\end{picture}
\end{center}
\vspace{0.3cm}
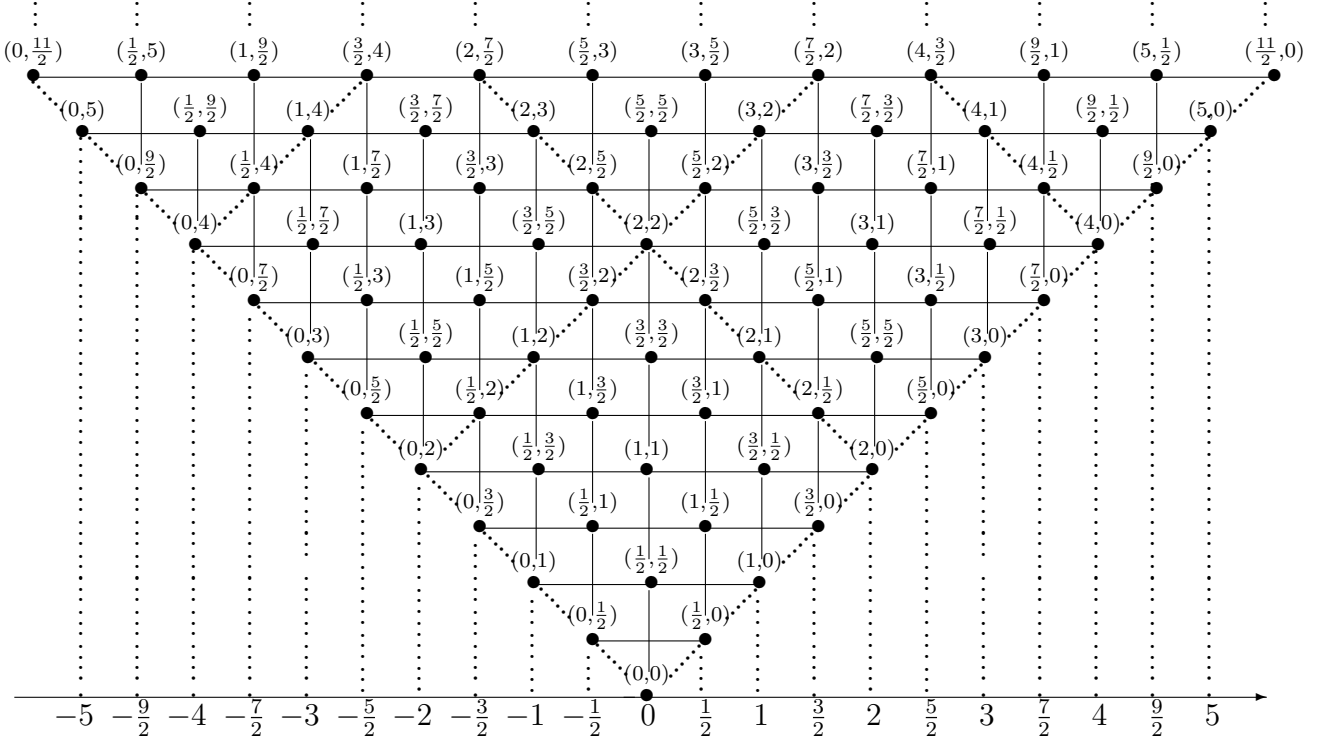
\begin{center}\begin{minipage}{32pc}{\small {\bf Fig.\,1:} The system of interlocking (cyclic) representations $(l,\dot{l})$ of the Lorentz group.}\end{minipage}\end{center}

In virtue of commutativity of the relations (\ref{Com2}) a space of an irreducible finite-dimensional representation of the group $\SL(2,\C)$ can be spanned on the totality of
$(2l+1)(2\dot{l}+1)$ basis ket-vectors $|l,m;\dot{l},\dot{m}\rangle$ and basis bra-vectors
$\langle l,m;\dot{l},\dot{m}|$, where $l,m,\dot{l},\dot{m}$ are integer
or half-integer numbers, $-l\leq m\leq l$, $-\dot{l}\leq
\dot{m}\leq \dot{l}$. Therefore,
\begin{eqnarray}
&&\sX_-|l,m;\dot{l},\dot{m}\rangle= \sqrt{(l+m)(l-m+1)}|l,m-1;\dot{l},\dot{m}\rangle
\;\;(m>-l),\nonumber\\
&&\sX_+|l,m;\dot{l},\dot{m}\rangle= \sqrt{(l-m)(l+m+1)}|l,m+1;\dot{l},\dot{m}\rangle
\;\;(m<l),\nonumber\\
&&\sX_3|l,m;\dot{l},\dot{m}\rangle=
m|l,m;\dot{l},\dot{m}\rangle,\nonumber\\
&&\langle l,m;\dot{l},\dot{m}|\sY_-=
\langle l,m;\dot{l},\dot{m}-1|\sqrt{(\dot{l}+\dot{m})(\dot{l}-\dot{m}+1)}\;\;(\dot{m}>-\dot{l}),\nonumber\\
&&\langle l,m;\dot{l},\dot{m}|\sY_+=
\langle l,m;\dot{l},\dot{m}+1|\sqrt{(\dot{l}-\dot{m})(\dot{l}+\dot{m}+1)}\;\;(\dot{m}<\dot{l}),\nonumber\\
&&\langle l,m;\dot{l},\dot{m}|\sY_3= \langle l,m;\dot{l},\dot{m}|\dot{m}.\label{Waerden}
\end{eqnarray}
In contrast to the
Gelfand-Naimark representation
for the Lorentz group \cite{GMS,Nai58},
which does not find a wide application in physics,
a representation (\ref{Waerden}) is a most useful in theoretical physics
(see, for example, \cite{AB,Sch61,RF,Ryd85}). This representation for the
Lorentz group was first given by van der Waerden in his brilliant book
\cite{Wa32}.
It should be noted here that the representation basis, defined by the
formulae (\ref{SL25})--(\ref{Waerden}), has an evident physical meaning.
For example, in the case of $(1,0)\oplus(0,1)$-representation space
there is an analogy with the photon spin states. Namely, the operators
$\sX$ and $\sY$ correspond to the right and left polarization states of the
photon. For that reason we will call the canonical basis consisting of the
vectors $\mid lm;\dot{l}\dot{m}\rangle$ as
{\it a helicity basis}.

Thus, a complex envelope of the group algebra $\mathfrak{sl}(2,\C)$, generating complex momentum, leads to a \textit{duality} which is mirrored in the appearance of the two spaces: a space of ket-vectors $|l,m;\dot{l},\dot{m}\rangle$ and a dual space of bra-vectors $\langle l,m;\dot{l},\dot{m}|$.
Further, equations for arbitrary spin chains (spin multiplets)
\[
\boldsymbol{\tau}_{l\dot{l}},\;\boldsymbol{\tau}_{l+\frac{1}{2},\dot{l}-\frac{1}{2}},\;
\boldsymbol{\tau}_{l+1,\dot{l}-1},\;\boldsymbol{\tau}_{l+\frac{3}{2},\dot{l}-\frac{3}{2}},\;\ldots,\;
\boldsymbol{\tau}_{\dot{l}l},
\]
(horizontal lines of the representation cone on the Fig.\,1) in the bivector space $\R^6$ have the form \cite{Var03,Var07}
\[
\sum^3_{j=1}\Lambda^{l\dot{l}}_j\frac{\partial\psi}{\partial a_j}-
i\sum^3_{j=1}\Lambda^{l\dot{l}}_j\frac{\partial\psi}{\partial a^\ast_j}+m^{(s)}\dot{\psi}=0,
\]
\[
\sum^3_{j=1}\Lambda^{l+\frac{1}{2},\dot{l}-\frac{1}{2}}_j\frac{\partial\psi}{\partial a_j}-
i\sum^3_{j=1}\Lambda^{l+\frac{1}{2}\dot{l}-\frac{1}{2}}_j\frac{\partial\psi}{\partial a^\ast_j}+m^{(s)}\dot{\psi}=0,
\]
\[
\sum^3_{j=1}\Lambda^{l+1,\dot{l}-1}_j\frac{\partial\psi}{\partial a_j}-
i\sum^3_{j=1}\Lambda^{l+1\dot{l}-1}_j\frac{\partial\psi}{\partial a^\ast_j}+m^{(s)}\dot{\psi}=0,
\]
\[
\sum^3_{j=1}\Lambda^{l+\frac{3}{2},\dot{l}-\frac{3}{2}}_j\frac{\partial\psi}{\partial a_j}-
i\sum^3_{j=1}\Lambda^{l+\frac{3}{2}\dot{l}-\frac{3}{2}}_j\frac{\partial\psi}{\partial a^\ast_j}+m^{(s)}\dot{\psi}=0,
\]
\[
\ldots\ldots\ldots\ldots\ldots\ldots\ldots\ldots\ldots\ldots\ldots\ldots
\]
\begin{equation}\label{BS2}
\sum^3_{j=1}\Lambda^{\dot{l}l}_j\frac{\partial\dot{\psi}}{\partial a_j}+
i\sum^3_{j=1}\Lambda^{\dot{l}l}_j\frac{\partial\dot{\psi}}{\partial a^\ast_j}+m^{(s)}\psi=0,
\end{equation}
where the spin\footnote{According to Weinberg theorem \cite{Wein}, a helicity
$\lambda$ of the particle, described by the representation
$\boldsymbol{\tau}_{l\dot{l}}$, is defined by an expression
$l-\dot{l}=\lambda$.} $s=l-\dot{l}$ changes as follows:
\[
l-\dot{l},\;l-\dot{l}+1,\;l-\dot{l}+2,\;l-\dot{l}+3,\,\ldots,\;\dot{l}-l.
\]
The system (\ref{BS2}) describes particle states with different masses and spins in $\R^6$. The state mass $m^{(s)}$, corresponding to the energy level $\sH_E\simeq\Sym_{(k,r)}$, is defined by the formula
\begin{equation}\label{MGY}
m^{(s)}=m_e\left(l+\frac{1}{2}\right)\left(\dot{l}+\frac{1}{2}\right),
\end{equation}
where $s=|l-\dot{l}|$. Mass spectrum and its dependence of the Lorentz group representations, characterizing by the pair $(l,\dot{l})$\footnote{Finite-dimensional representations $\boldsymbol{\tau}_{l\dot{l}}$ of the Lorentz group (usually denoted as $\fD^{(j/2,r/2)}$, here $j/2=l$, $r/2=\dot{l}$) play an important role in the axiomatic quantum field theory \cite{SW64,BLOT}. For more details about interlocking representations $\boldsymbol{\tau}_{l\dot{l}}$ see \cite{GY48,AD72,PS83}.}, was given in \cite{Var15}. Usually one consider the mass spectrum dependence of spin $s$ only (Lorentz group representation is fixed). As it is known, in this case non-physical mass spectrum arises, $m_i\sim 1/s_i$, that is, for very large $s_i$, masses are very small (the first mass formula of this type was given by Majorana \cite{Maj32} and similar mass formulas were  considered by Gelfand and Yaglom \cite{GY48}, see also \cite{Esp12,Ulr13}). In the paper \cite{Var15} it has been shown that for representations $(l,\dot{l})$ of the Lorentz group the mass is proportional to $(l+1/2)(\dot{l}+1/2)$. Hence it immediately follows that particles (more precisely, particle states) with the \textit{same} spin but \textit{distinct} masses are described by \textit{different} representations of Lorentz group.

The operators $\boldsymbol{\Lambda}=(\Lambda^{l\dot{l}}_1,\Lambda^{l\dot{l}}_2,\Lambda^{l\dot{l}}_3)$ and $\overset{\ast}{\boldsymbol{\Lambda}}=(\Lambda^{\dot{l}l}_1,\Lambda^{\dot{l}l}_2,\Lambda^{\dot{l}l}_3)$ in (\ref{BS2}) have a common system of eigenfunctions with the energy operator $H$ (and operators $\sX_l$, $\sY_l$ of the complex momentum) for the each spin value $s=l-\dot{l}$. Hence it follows that $\boldsymbol{\Lambda}$ and $\overset{\ast}{\boldsymbol{\Lambda}}$ are Hermitian operators. It has been shown \cite{Var16} that a structure $\boldsymbol{\Lambda}$, $\overset{\ast}{\boldsymbol{\Lambda}}$ depends on the structure of elementary divisors.
In the case of $(l,0)\oplus(0,\dot{l})$ all elementary divisors of the operators $\boldsymbol{\Lambda}$ and $\overset{\ast}{\boldsymbol{\Lambda}}$ are simple, and Jordan form of the matrices of $\boldsymbol{\Lambda}$ and $\overset{\ast}{\boldsymbol{\Lambda}}$ is a diagonal matrix
\begin{equation}\label{SimpleJ}
\bJ(\Lambda)=-\bJ(\overset{\ast}{\Lambda})=\text{{\rm diag}}\left(\lambda_1,\lambda_2,\ldots,\lambda_{2l+1}\right),
\end{equation}
where $\lambda_1,\lambda_2,\ldots,\lambda_{2l+1}$ are eigenvalues of the operators $\boldsymbol{\Lambda}$ and $\overset{\ast}{\boldsymbol{\Lambda}}$.

In the case of $(l,\dot{l})\oplus(\dot{l},l)$ the operators $\boldsymbol{\Lambda}$ and $\overset{\ast}{\boldsymbol{\Lambda}}$ have elementary divisors with multiple roots, and Jordan form of the matrices of $\boldsymbol{\Lambda}$ and $\overset{\ast}{\boldsymbol{\Lambda}}$ is a block-diagonal matrix. For example, characteristic polynomial of $\Lambda^{s\frac{k}{2}}_3$ is
\begin{multline}
\boldsymbol{\Delta}(\Lambda^{s\frac{k}{2}}_3)=(\lambda-sk/2)^{m_1}(\lambda-(1-s)k/2)^{m_2}(\lambda-(2-s)k/2)^{m_3}
\ldots(\lambda-1/2)^{m_p}\lambda^{2s}\times\\
\times(\lambda+sk/2)^{m_{-1}}(\lambda+(1-s)k/2)^{m_{-2}}(\lambda+(2-s)k/2)^{m_{-3}}\ldots(\lambda+1/2)^{m_{-p}},
\nonumber
\end{multline}
where
\[
2\leq m_1\leq m^1_{\text{{\rm max}}},
\]
\[
2\leq m_2\leq m^2_{\text{{\rm max}}},
\]
\[
\ldots\ldots\ldots\ldots\ldots
\]
\[
2\leq m_{-p}\leq m^{-p}_{\text{{\rm max}}},
\]
here $m^1_{\text{max}}$ is a number of products among $(2s+1)(k+1)$ products of the $sk$-basis, which equal to eigenvalue $sk/2$; $m^2_{\text{max}}$ is a number of products among $(2s+1)(k+1)$ products of the $sk$-basis, which equal to eigenvalue $(1-s)k/2$ and so on.

The Jordan form of $\Lambda^{s\frac{k}{2}}_3$ is
\[
\bJ(\Lambda^{s\frac{k}{2}}_3)=\text{{\rm diag}}\left(\bJ({}_{s\frac{k}{2}}\Lambda^{s\frac{k}{2}}_3), \bJ({}_{\frac{(1-s)k}{2}}\Lambda^{s\frac{k}{2}}_3),\ldots,\bJ({}_{\frac{1}{2}}\Lambda^{s\frac{k}{2}}_3),
\bJ({}_{0}\Lambda^{s\frac{k}{2}}_3),\bJ({}_{-\frac{1}{2}}\Lambda^{s\frac{k}{2}}_3),\ldots,
\bJ({}_{-\frac{sk}{2}}\Lambda^{s\frac{k}{2}}_3)\right),
\]
where
\[
\bJ({}_{\frac{sk}{2}}\Lambda^{s\frac{k}{2}}_3)=\begin{Vmatrix}
\frac{sk}{2} & 1 & 0 & \dots & 0\\
0 & \frac{sk}{2} & 1 & \dots & 0\\
.&&\hdotsfor[3]{3}\\
.&&&\hdotsfor[3]{2}\\
.&&&.&1\\
0 & 0 & 0 & \dots & \frac{sk}{2}
\end{Vmatrix}
\]
is a Jordan cell corresponding to elementary divisor $(\lambda-sk/2)^{m_1}$ and so on. Thus, the states (cyclic representations) of the form $(l,0)\oplus(0,\dot{l})$ have a trivial Jordan structure that corresponds to a spectrum of ``point-like'' (structureless) particle. Whereas the states of the form $(l,\dot{l})\oplus(\dot{l},l)$ have non-trivial Jordan structure, and corresponding cyclic representations contain invariant subspaces (there are elementary divisors of higher order). In this case we come to a spectrum of inhomogeneous (``composite'') particle which endowed by a grain (parton) structure.

\subsection{Matter spectrum and Gelfand-Neumark-Segal construction}
It is well-known that a ground of algebraic formulation of quantum theory is the Gelfand-Neumark-Segal construction (GNS), defined by a canonical correspondence $\omega\leftrightarrow\pi_\omega$ between states and cyclic representations of $C^\ast$-algebra \cite{Emh,BLOT,Hor86}. One of the most important aspects in theory of $C^\ast$-algebras is a duality between states and representations. A relation between states and irreducible representations of operator algebras was first formulated by Segal \cite{Se47}. Let $\pi$ be a some representation of the algebra $\fA$ in the Hilbert space $\sH_\infty$, then for any non-null vector $\left|\Phi\right\rangle\in\sH_\infty$ the expression
\begin{equation}\label{VectState}
\omega_\Phi(A)=\frac{\langle\Phi\mid\pi(A)\Phi\rangle}{\langle\Phi\mid\Phi\rangle}
\end{equation}
defines a state $\omega_\Phi(A)$ of the algebra $\fA$. $\omega_\Phi(A)$ is called a \textit{vector state} associated with the representation $\pi$ ($\omega_\Phi(A)$ corresponds to the vector $\left|\Phi\right\rangle$). Let $S_\pi$ be a set of all states associated with the representation $\pi$. Two representations $\pi_1$ and $\pi_2$ with one and the same set of associated states (that is, $S_{\pi_1}=S_{\pi_2}$) are called phenomenologically equivalent sets (it corresponds to unitary equivalent representations). Moreover, the set $PS(\fA)$ of the all pure states of $C^\ast$-algebra $\fA$ coincides with the set of all vector states associated with the all \textit{irreducible representations} of the algebra $\fA$.

Further, let $\pi$ be a representation of $C^\ast$-algebra $\fA$ in $\sH_\infty$ and let $\left|\Phi\right\rangle$ be a \textit{cyclic vector}\footnote{Vector $\left|\Phi\right\rangle\in\sH_\infty$ is called a \textit{cyclic vector} for the representation $\pi$, if the all vectors $\left|\pi(A)\Phi\right\rangle$ (where $A\in\fA$) form a total set in $\sH_\infty$, that is, such a set, for which a closing of linear envelope is dense everywhere in $\sH_\infty$. $\pi$ with the cyclic vector is called a \textit{cyclic representation}.} of the representation $\pi$ defining the state $\omega_\Phi$. In accordance with Gelfand-Neumark-Segal construction (see \cite{BLOT}) the each state defines a some representation of the algebra $\fA$. At this point, resulting representation is irreducible exactly when the state is pure. Close relationship between states and representations of $C^\ast$-algebra, based on the GNS construction, allows us to consider representations of the algebra as an effective tool for organization of states. This fact becomes more evident at the concrete realization $\pi(\fA)$.

Let us consider a concrete realization of the operator algebra $\fA$. A transition $\fA\Rightarrow\pi(\fA)$ from $\fA$ to a concrete algebra $\pi(\fA)$ is called sometimes as `clothing'. So, the basic observable is an \textit{energy} which represented by a Hermitian operator $H$. Let $G=\SO_0(1,3)\simeq\SL(2,\C)/\dZ_2$ be the group of fundamental symmetry, where $\SO_0(1,3)$ is the Lorentz group. Let $\widetilde{G}\simeq\SL(2,\C)$ be the \textit{universal covering} of $\SO_0(1,3)$. Let $H$ be the energy operator defined on the separable Hilbert space $\sH_\infty$. Then all the possible values of energy (states) are eigenvalues of the operator $H$. At this point, if $E_1\neq E_2$ are eigenvalues of $H$, and $\left|\Phi_1\right\rangle$ and $\left|\Phi_2\right\rangle$ are corresponding eigenvectors in the space $\sH_\infty$, then $\langle\Phi_1\mid\Phi_2\rangle=0$. All the eigenvectors, belonging to a given eigenvalue $E$, form (together with the null vector) an \textit{eigenvector subspace} $\sH_E$ of the Hilbert space $\sH_\infty$. All the eigenvector subspaces $\sH_E\in\sH_\infty$ are finite-dimensional. A dimensionality $r$ of $\sH_E$ is called a \textit{multiplicity} of the eigenvalue $E$. When $r>1$ the eigenvalue $E$ is \textit{$r$-fold degenerate}. Further, let $\sX_l$, $\sY_l$ be infinitesimal operators of the complex envelope of the group algebra $\mathfrak{sl}(2,\C)$ for the universal covering $\widetilde{G}$, $l=1,2,3$. As is known \cite{BHJ26}, the energy operator $H$ commutes with the all operators in $\sH_\infty$, which represent a Lie algebra of the group $\widetilde{G}$. Let us consider an arbitrary eigenvector subspace $\sH_E$ of the energy operator $H$. Since the operators $\sX_l$, $\sY_l$ and $H$ commute with the each other, then, as is known \cite{Dir}, for these operators we can build a common system of eigenfunctions. It means that the subspace $\sH_E$ is invariant with respect to operators $\sX_l$, $\sY_l$ (moreover, the operators $\sX_l$, $\sY_l$ can be considered \textit{only on} $\sH_E$). Further, we suppose that there is a some \textit{local representation} of the group $\widetilde{G}$ defined by the operators acting in the space $\sH_\infty$. At this point, we assume that all the representing operators commute with $H$. Then the each eigenvector subspace $\sH_E$ of the energy operator is invariant with respect to operators of complex momentum $\sX_l$, $\sY_l$. It allows us to identify subspaces $\sH_E$ with symmetrical spaces $\Sym_{(k,r)}$ of interlocking representations $\boldsymbol{\tau}_{k/2,r/2}$ of the Lorentz group. Thus, we obtain a concrete realization (`clothing') of the operator algebra $\pi(\fA)\rightarrow\pi(H)$, where $\pi\equiv\boldsymbol{\tau}_{k/2,r/2}$. The system of interlocking representations of the Lorentz group is shown on the Fig.\,1 (for more details see \cite{Var03,Var07}). Hence it follows that the each possible value of energy (energy level) is a vector state of the form (\ref{VectState}):
\begin{equation}\label{VectState2}
\omega_\Phi(H)=\frac{\langle\Phi\mid\pi(H)\Phi\rangle}{\langle\Phi\mid\Phi\rangle}=
\frac{\langle\Phi\mid\boldsymbol{\tau}_{k/2,r/2}(H)\Phi\rangle}{\langle\Phi\mid\Phi\rangle},
\end{equation}
The state $\omega_\Phi(H)$ is associated with the representation $\pi\equiv\boldsymbol{\tau}_{k/2,r/2}$ and the each $\omega_\Phi(H)$ corresponds to a non-null (cyclic) vector $\left|\Phi\right\rangle\in\sH_\infty$.

Further, in virtue of the isomorphism $\SL(2,\C)\simeq\spin_+(1,3)$ we will consider the universal covering $\widetilde{G}$ as a \textit{spinor group}. It allows us to associate in addition a \textit{spinor structure} with the each cyclic vector $\left|\Phi\right\rangle\in\sH_\infty$ (in some sense, it be a second layer in `clothing' of the operator algebra). Spintensor representations of the group $\widetilde{G}\simeq\spin_+(1,3)$ form a \textit{substrate} of interlocking representations $\boldsymbol{\tau}_{k/2,r/2}$ of the Lorentz group realized in the spaces $\Sym_{(k,r)}\subset\dS_{2^{k+r}}$, where $\dS_{2^{k+r}}$ is a spinspace. In its turn, as it is known \cite{Lou91}, a spinspace is a minimal left ideal of the Clifford algebra $\cl_{p,q}$, that is, there exists an isomorphism $\dS_{2^m}(\K)\simeq I_{p,q}=\cl_{p,q}f$, where $f$ is a primitive idempotent of $\cl_{p,q}$, and $\K=f\cl_{p,q}f$ is a division ring of the algebra $\cl_{p,q}$, $m=(p+q)/2$. The complex spinspace $\dS_{2^m}(\C)$ is a complexification $\C\otimes I_{p,q}$ of the minimal left ideal $I_{p,q}$ of the real subalgebra $\cl_{p,q}$. So, $\dS_{2^{k+r}}(\C)$ is the minimal left ideal of the complex algebra $\C_{2k}\otimes\overset{\ast}{\C}_{2r}\simeq\C_{2(k+r)}$ (for more details see \cite{Var15,Var16}). Let us define a system of \textit{basic cyclic vectors} endowed with the complex spinor structure (these vectors correspond to the system of interlocking representations of the Lorentz group):
\begin{eqnarray}
&&\mid\C_0,\boldsymbol{\tau}_{0,0}(H)\Phi\rangle;\nonumber\\
&&\mid\C_2,\boldsymbol{\tau}_{1/2,0}(H)\Phi\rangle,\quad\mid\overset{\ast}{\C}_2,\boldsymbol{\tau}_{0,1/2}(H)\Phi\rangle;
\nonumber\\
&&\mid\C_2\otimes\C_2,\boldsymbol{\tau}_{1,0}(H)\Phi\rangle,\quad
\mid\C_2\otimes\overset{\ast}{\C}_2,\boldsymbol{\tau}_{1/2,1/2}(H)\Phi\rangle,\quad
\mid\overset{\ast}{\C}_2\otimes\overset{\ast}{\C}_2,\boldsymbol{\tau}_{0,1}(H)\Phi\rangle;\nonumber\\
&&\mid\C_2\otimes\C_2\otimes\C_2,\boldsymbol{\tau}_{3/2,0}(H)\Phi\rangle,\quad
\mid\C_2\otimes\C_2\otimes\overset{\ast}{\C}_2,\boldsymbol{\tau}_{1,1/2}(H)\Phi\rangle,\quad
\mid\C_2\otimes\overset{\ast}{\C}_2\otimes\overset{\ast}{\C}_2,\boldsymbol{\tau}_{1/2,1}(H)\Phi\rangle,\nonumber\\
&&\mid\overset{\ast}{\C}_2\otimes\overset{\ast}{\C}_2\otimes\overset{\ast}{\C}_2,\boldsymbol{\tau}_{0,3/2}(H)\Phi\rangle;
\nonumber\\
&&\ldots\ldots\ldots\ldots\ldots\ldots\ldots\ldots\ldots\ldots\ldots\ldots\ldots\ldots\ldots\ldots\ldots\ldots\ldots
\nonumber
\end{eqnarray}
Therefore, in accordance with GNS construction we have complex vector states of the form
\begin{equation}\label{VectState3}
\omega^c_\Phi(H)=
\frac{\langle\Phi\mid\C_{2(k+r)},\boldsymbol{\tau}_{k/2,r/2}(H)\Phi\rangle}{\langle\Phi\mid\Phi\rangle},
\end{equation}
The states $\omega^c_\Phi(H)$ are associated with the complex representations $\boldsymbol{\tau}_{k/2,r/2}(H)$ and cyclic vectors $\left|\Phi\right\rangle\in\sH_\infty$.

As is known, in the Lagrangian formalism of the standard (local) quantum field theory \textit{charged particles} are described by \textit{complex fields}. In our case, pure states of the form (\ref{VectState3}) correspond to \textit{charged states}. At this point, the sign of charge is changed under action of the pseudoautomorphism $\cA\rightarrow\overline{\cA}$ of the complex spinor structure (for more details see \cite{Var01a,Var04,Var14}). Following to analogy with the Lagrangian formalism, where \textit{neutral particles} are described by \textit{real fields}, we introduce vector states of the form
\begin{equation}\label{VectState4}
\omega^r_\Phi(H)=
\frac{\langle\Phi\mid\cl_{p,q},\boldsymbol{\tau}_{k/2,r/2}(H)\Phi\rangle}{\langle\Phi\mid\Phi\rangle}.
\end{equation}
The states (\ref{VectState4}) are associated with the real representations $\boldsymbol{\tau}_{k/2,r/2}(H)$, that is, these representations are endowed with a \textit{real spinor structure}, where $\cl_{p,q}$ is a real subalgebra of $\C_{2(k+r)}$. States of the form (\ref{VectState4}) correspond to \textit{neutral states}. Since the real spinor structure is appeared in the result of reduction $\C_{2(k+r)}\rightarrow\cl_{p,q}$, then (as a consequence) a \textit{charge conjugation} $C$ (pseudoautomorphism $\cA\rightarrow\overline{\cA}$) for the algebras $\cl_{p,q}$ over the real number field $\F=\R$ and quaternionic division ring $\K\simeq\BH$ (the types $p-q\equiv 4,6\pmod{8}$) is reduced to \textit{particle-antiparticle interchange} $C^\prime$ (see \cite{Var01a,Var04,Var14}). As is known, there exist two classes of neutral particles: 1) particles which have antiparticles, such as neutrons, neutrino\footnote{However, it should be noted that the question whether neutrinos are Dirac or Majorana particles (truly neutral fermions) is still open (the last hypothesis being preferred by particle physicists).} and so on; 2) particles which coincide with their antiparticles (for example, photons, $\pi^0$-mesons and so on), that is, so-called \textit{truly neutral particles}. The first class is described by neutral states $\omega^r_\Phi(H)$ with the algebras $\cl_{p,q}$ over the field $\F=\R$ with the rings $\K\simeq\BH$ and $\K\simeq\BH\oplus\BH$ (types $p-q\equiv 4,6\pmod{8}$ and $p-q\equiv 5\pmod{8}$). With the aim to describe the second class of neutral particles we introduce \textit{truly neutral states} $\omega^{r_0}_\Phi(H)$ with the algebras $\cl_{p,q}$ over the number field $\F=\R$ and real division rings $\K\simeq\R$ and $\K\simeq\R\oplus\R$ (types $p-q\equiv 0,2\pmod{8}$ and $p-q\equiv 1\pmod{8}$). In the case of states $\omega^{r_0}_\Phi(H)$ pseudoautomorphism $\cA\rightarrow\overline{\cA}$ is reduced to identical transformation (particle coincides with its antiparticle).

In \cite{Var16b} it has been shown that basic energetic levels (states) of matter spectrum are constructed in terms of cyclic representations within GNS construction. A concrete realization of the operator algebra (energy operator) is defined via the spinor structure (SS). SS-morphisms define charge and discrete symmetries (charge conjugation $C$, parity transformation $P$, time reversal $T$ and their combinations) \cite{Var99,Var01a,Var01,Var05,Var05a,Var11,Var14}, associated with the each cyclic representation.
\subsection{Physical Hilbert space}
A set of pure states $\omega_\Phi(H)$, defined according to GNS construction by the equality (\ref{VectState2}), at the execution of condition $\omega_\Phi(H)\geq 0$ forms a \textit{physical Hilbert space}
\[
\bsH_{\rm phys}=\bsH^S\otimes\bsH^Q\otimes\bsH_\infty.
\]
It is easy to verify that axioms of addition, multiplication and scalar (inner) product are fulfilled for the vectors $\omega_\Phi(H)\rightarrow\left|\Psi\right\rangle\in\bsH_{\rm phys}$. We assume that a so-defined Hilbert space is \textit{nonseparable}, that is, in general case the axiom of separability is not executed in $\bsH_{\rm phys}$.

The space $\bsH_{\rm phys}$ is a second member of the pair $(\sH_\infty,\bsH_{\rm phys})$\footnote{In accordance with Heisenberg-Fock conception \cite{Heisen,Fock}, reality has a two-level structure: \textit{potential reality} and \textit{actual reality}. Heisenberg claimed that any quantum object (for example, elementary particle) belongs to both sides of reality: at first, to potential reality as a superposition, and to actual reality after reduction of superposition (for more details see \cite{Heisen,Fock,Var15b}).}, which defines two-level structure of the Hilbert space of single quantum system (quantum domain). Therefore, $\bsH_{\rm phys}$ describes spin and charge degrees of freedom of the particle state. In accordance with the charge degrees of freedom we separate three \textit{basic subspaces} in $\bsH_{\rm phys}$.\\
1) \textit{Subspace of charged states} $\bsH^\pm_{\rm phys}=\bsH^S\otimes\bsH^\pm\otimes\bsH_\infty$.\\
2) \textit{Subspace of neutral states} $\bsH^0_{\rm phys}=\bsH^S\otimes\bsH^0\otimes\bsH_\infty$.\\
3) \textit{Subspace of truly neutral states} $\bsH^{\overline{0}}_{\rm phys}=\bsH^S\otimes\bsH^{\overline{0}}\otimes\bsH_\infty$.\\
Basis vectors $\left|\Psi\right\rangle\in\bsH^\pm_{\rm phys}$ are formed by the states $\omega^c_\Phi(H)$ (see (\ref{VectState3})). Correspondingly, $\left|\Psi\right\rangle\in\bsH^0_{\rm phys}$ and $\left|\Psi\right\rangle\in\bsH^{\overline{0}}_{\rm phys}$ are formed by the states $\omega^{r}_\rho(H)$ and $\omega^{r_0}_\rho(H)$.

Pure states (cyclic representations) are divided with respect to a charge (an action of SS-pseudoautomorphism) on the subsets of charged, neutral and truly neutral states. The structure of matter spectrum is defined by a partition of a physical Hilbert space (state space) on the coherent subspaces \cite{Var16b}
\begin{equation}\label{Decomp3}
\bsH_{\rm phys}=\bigoplus_{b,\ell\in\dZ}\left[\bsH^\pm_{\rm phys}(b,\ell)\bigoplus\bsH^0_{\rm phys}(b,\ell)\bigoplus
\bsH^{\overline{0}}_{\rm phys}(b,\ell)\right],
\end{equation}
where
\[
\bsH^Q_{\rm phys}(b,\ell)=\bigoplus^{|l-\dot{l}|}_{s=-|l-\dot{l}|}\bsH^{2|s|+1}\otimes\bsH^Q(b,\ell)\otimes\bsH_\infty,\quad
Q=\{\pm,0,\overline{0}\}.
\]
Here $b$ and $\ell$ are eigenvalues of baryon $B$ and lepton $L$ charges. At the values of electric charge $Q=\{\pm,0,\overline{0}\}$ we have three basic subspaces: subspace of charged states $\bsH^\pm_{\rm phys}(b,\ell)$, subspace of neutral states $\bsH^0_{\rm phys}(b,\ell)$ and subspace of truly neutral states $\bsH^{\overline{0}}_{\rm phys}(b,\ell)$.
\section{Lepton sector}
The lepton sector contains states belonging to coherent subspaces $\bsH^\pm_{\rm phys}(0,\ell)$ and $\bsH^0_{\rm phys}(0,\ell)$\footnote{Or $\bsH^{\overline{0}}_{\rm phys}(0,\ell)$ in case of Majorana neutrinos (truly neutral fermions).}, where the lepton number (flavor) $\ell$ takes three values: $\ell_\alpha=\{\ell_e,\ell_\mu,\ell_\tau\}$. All the states (particles) of the lepton sector are fermions of the spin 1/2, therefore, according to GNS construction, they are described by cyclic representations belonging to spin lines 1/2 and -1/2 (see Fig.\,1). For the charged leptons we have the following coherent subspaces:
\begin{equation}\label{ChargeLep-}
\bsH^\pm_{\rm phys}(0,\ell_\alpha)=\bsH^2\otimes\bsH^\pm(0,\ell_\alpha)\otimes\bsH_\infty,
\end{equation}
\begin{equation}\label{ChargeLep+}
\overset{\ast}{\bsH}{}^\mp_{\rm phys}(0,\ell_\alpha)=\bsH^2\otimes\bsH^\mp(0,\ell_\alpha)\otimes\bsH_\infty,
\end{equation}
Charged leptons $e^-$, $\mu^-$, $\tau^-$ and their antiparticles ($e^+$, $\mu^+$, $\tau^+$) form qubit-like nonseparable states in the spaces (\ref{ChargeLep-}) and (\ref{ChargeLep+}), respectively.

Using the formula (\ref{MGY}), we define now masses of the charged leptons. The least rest mass corresponds to electron:
\[
m_e=0,511\;\;\text{MeV}.
\]
This experimental value of electron mass is a fundamental constant in the formula (\ref{MGY}). In other words, $m_e$ is an original point in the mass frame of reference, and theoretical masses of all other states (``elementary particles'') are calculated from which. The following (on the mass scale) lepton is a muon $\mu^-$ with experimental mass value
\begin{equation}\label{Mexp}
m_\mu=105,66\;\;\text{MeV}.
\end{equation}
The muon mass is approximately 207 times (206,77) as much the electron mass. This mass value corresponds to cyclic representation $(l,\dot{l})$ on the spin-1/2 line at $l=14$ and $\dot{l}=27/2$:
\begin{equation}\label{Mtheor}
m^{\frac{1}{2}}_\mu=m_e\left(l+\frac{1}{2}\right)\left(\dot{l}+\frac{1}{2}\right)=103,73.
\end{equation}
The absolute error between experimental (\ref{Mexp}) and calculated (\ref{Mtheor}) values is equal to $-1,93$; the relative error is $-1,98\%$.

Analogously, for the $\tau$-lepton with experimental mass value
\begin{equation}\label{Texp}
m_\tau=1776,84\;\;\text{MeV}
\end{equation}
we have a cyclic representation $(l,\dot{l})$ on the spin-1/2 line at $l=117/2$, $\dot{l}=58$ with the mass
\begin{equation}\label{Ttheor}
m^{\frac{1}{2}}_\tau=m_e\left(l+\frac{1}{2}\right)\left(\dot{l}+\frac{1}{2}\right)=1763,72.
\end{equation}
The absolute error between (\ref{Texp}) and (\ref{Ttheor}) is $-13,12$; the relative error is $-0,74\%$\footnote{In relation with the masses of charged leptons to be of interest to recall a further empirical mass formula. In 1983, Koide \cite{Koi83} proposed the following formula:
\[
m_e+m_\mu+m_\tau=\frac{2}{3}\left(\sqrt{m_e}+\sqrt{m_\mu}+\sqrt{m_\tau}\right)^2,
\]
which relates the masses of charged leptons. It should be noted that Koide predicted (on the ground of this formula) the mass $\tau$-lepton (1777 MeV) which coincides very exactly with the modern mass value of this particle (see (\ref{Texp})), whereas in 1983 this value had 1784 Mev. About Koide's formula and its interpretations and applications see \cite{Gsp05,Foot,Esp}.}.

Neutrino states of matter spectrum are described within coherent subspaces $\bsH^0_{\rm phys}(0,\ell)$ (Dirac neutrinos) or $\bsH^{\overline{0}}_{\rm phys}(0,\ell)$ (Majorana neutrinos). All neutrino states are neutral (or truly neutral) fermions of the spin 1/2. According to GNS construction, these states define cyclic representations of the spin lines 1/2 and -1/2. For the Dirac neutrinos we have two copies of coherent subspaces
\begin{equation}\label{Neutrino1}
\bsH^0_{\rm phys}(0,\nu_\alpha)=\bsH^2\otimes\bsH^0(0,\nu_\alpha)\otimes\bsH_\infty,
\end{equation}
\begin{equation}\label{Neutrino2}
\overline{\bsH}^0_{\rm phys}(0,\nu_\alpha)=\bsH^2\otimes\overline{\bsH}^0(0,\nu_\alpha)\otimes\bsH_\infty.
\end{equation}
These subspaces are transformed into each other under action of the SS-pseudoautomorphism. In case of Majorana neutrinos, for the each neutrino flavor ($\nu_\alpha=\{\nu_e,\nu_\mu,\nu_\tau\}$) we have the only one copy of coherent subspace
\begin{equation}\label{Neutrino3}
\bsH^{\overline{0}}_{\rm phys}(0,\nu_\alpha)=\bsH^2\otimes\bsH^{\overline{0}}(0,\nu_\alpha)\otimes\bsH_\infty.
\end{equation}
In this case SS-pseodoautomorphism is reduced to an identical transformation.

Further, for any kind of neutrino (Dirac or Majorana) all the neutrino states are defined by quantum superpositions
\begin{equation}\label{Neutrino}
\nu_\alpha=\sum_iU_{\alpha i}\nu_i,
\end{equation}
where $\nu_i$ are neutrino mass states, $U_{\alpha i}$ are elements of the Pontecorvo-Maki-Nakagawa-Sakata mixing matrix (PMNS)
\begin{multline}
U=\begin{pmatrix} 1 & 0 & 0\\
0 & \cos\theta_{23} & \sin\theta_{23}\\
0 & -\sin\theta_{23} & \cos\theta_{23}
\end{pmatrix}\times\\
\times\begin{pmatrix} \cos\theta_{13} & 0 & \sin\theta_{13}\exp(-i\delta)\\
0 & 1 & 0\\
-\sin\theta_{13}\exp(-i\delta) & 0 & \cos\theta_{13}
\end{pmatrix}\times\\
\times\begin{pmatrix} \cos\theta_{12} & \sin\theta_{12} & 0\\
-\sin\theta_{12} & \cos\theta_{12} & 0\\
0 & 0 & 1
\end{pmatrix}.\nonumber
\end{multline}
Here the mixing angle $\theta_{12}$ describes ``solar'' and ``reactor'' oscillations, and $\theta_{23}$ serves for description of ``atmospheric'' and ``acceleration'' oscillations \cite{Olsh}. Thus, all neutrinos form superpositions of the type (\ref{Neutrino}) in the coherent subspaces (\ref{Neutrino1})--(\ref{Neutrino2}) or (\ref{Neutrino3}).

Neutrino oscillations show that lepton number (flavor) is not conserved for the neutrino, that is, neutrino with a definite flavor cannot be considered as a particle with a definite mass, therefore, it presents by itself a quantum superposition of massive neutrino states\footnote{It should be noted that mixing matrix $U$ is attributed to neutrino only for convenience. In fact, $U$ is a general lepton mixing matrix \cite{Bed}. Can we expect that instead charged lepton with a definite flavor there exists a quantum superposition $\ell_i=\sum_\alpha U_{\alpha i}\ell_\alpha$ and, as a consequence, there are oscillations of charged leptons \cite{Bed}? It is natural to assume that the superposition structure (\ref{Neutrino}), which is characteristic of neutrino, is also a general feature of the all objects of microworld. To all appearances, an absence of experimental evidences in favour of existence of oscillations of charged leptons consists in the fact that for majority of practical cases the mass square difference of charged leptons is too much for a creation of charged leptons in a coherent quantum superposition.}.

\section{Meson sector}
A meson sector of the matter spectrum is formed by the states belonging to coherent subspaces $\bsH^\pm_{\rm phys}(0,0)$ and $\bsH^0_{\rm phys}(0,0)$ (or $\bsH^{\overline{0}}_{\rm phys}(0,0)$), where the lepton $\ell$ and baryon $b$ numbers are equal to zero. All mesons are states with integer spin. Therefore, in accordance with GNS construction all meson states are described by cyclic representations belonging to spin lines of integer spin with the non-null mass. Below we give tables for the lines of spin $0$, $1$, $\ldots$, $4$, including all known at the present time (according to Particle Data Group \cite{PDG}\footnote{In the tables we include only states with a three-star status (at the minimum).}) mesons of these spin lines. In the first column of the each table we give a standard designation of the state and its mass in MeV. The second column of the table contains a calculated (according to (\ref{MGY})) mass value of the corresponding state. The third column contains a relative error between experimental and calculated values. In the fourth column we give parameters $l$ and $\dot{l}$ of the corresponding cyclic representation.
\begin{center}
{\textbf{Table\,I.} Mesons: spin-0 line.}
{\renewcommand{\arraystretch}{1}
\begin{tabular}{|c||l|l|c|l|}\hline
 & State and mass (exp.) -- MeV & Mass (theor.) & Error \% & $(l,\dot{l})$\\ \hline\hline
1. & $\pi$ -- $139,57$ & $139,12$ & $-0,32$ & $(16,16)$\\
2. & $K$ -- $493,67$ & $491,07$ & $-0,52$ & $(61/2,61/2)$\\
3. & $f_0(500)$ & $507,03$ & $+1,40$ & $(31,31)$\\
4. & $\eta$ -- $547,86$ & $539,74$ & $-1,46$ & $(32,32)$\\
5. & $\eta^\prime(958)$ -- $957,78$ & $966,94$ & $+0,95$ & $(43,43)$\\
6. & $a_0(980)$ & $989,29$ & $+0,94$ & $(87/2,87/2)$\\
7. & $f_0(980)$ -- $990$ & $989,29$ & $-0,07$ & $(87/2,87/2)$\\
8. & $\eta(1295)$ & $1277,50$ & $-1,35$ & $(99/2,99/2)$\\
9. & $\pi(1300)$ & $1303,18$ & $+0,24$ & $(50,50)$\\
10. & $f_0(1370)$ & $1381,74$ & $+0,85$ & $(103/2,103/2)$\\
11. & $\eta(1405)$ -- $1408,8$ & $1408,44$ & $-0,02$ & $(52,52)$\\
12. & $K^\ast_0(1430)$ & $1435,39$ & $+0,37$ & $(105/2,105/2)$\\
13. & $a_0(1450)$ & $1435,39$ & $-1,00$ & $(105/2,105/2)$\\
14. & $\eta(1475)$ & $1462,61$ & $-0,84$ & $(53,53)$\\
15. & $f_0(1500)$ & $1490,08$ & $-0,66$ & $(107/2,107/2)$\\
16. & $f_0(1710)$ -- $1723\pm{}^{6}_{5}$ & $1719$ & $-0,23$ & $(115/2,115/2)$\\
17. & $\pi(1800)$ -- $1812\pm 12$ & $1809,07$ & $-0,16$ & $(59,59)$\\
18. & $D$ -- $1869,62$ & $1870,39$ & $+0,04$ & $(60,60)$\\
19. & $D^\pm_s$ -- $1968,30$ & $1964,28$ & $-0,21$ & $(123/2,123/2)$\\
20. & $D^\ast_{s0}(2317)^\pm$ -- $2317,7\pm 0,6$ & $2328,24$ & $+0,48$ & $(67,67)$\\
21. & $\eta_c(1S)$ -- $2983,6\pm 0,61$ & $2990,50$ & $+0,23$ & $(76,76)$\\
22. & $\chi_{c0}(1P)$ -- $3414,75$ & $3394,19$ & $-0,60$ & $(81,81)$\\
\hline
\end{tabular}
}
\end{center}
\begin{center}
{\renewcommand{\arraystretch}{1.0}
\begin{tabular}{|c||l|l|c|l|}\hline
& State and mass (exp.) -- MeV & Mass (theor.) & Error \% & $(l,\dot{l})$\\ \hline\hline
23. & $\eta_c(2S)$ -- $3639,2\pm 1,2$ & $3648,67$ & $+0,26$ & $(84,84)$\\
24. & $\chi_{c0}(3915)$ -- $3918,4\pm 1,9$ & $3912,34$ & $-0,15$ & $(87,87)$\\
25. & $B$ -- $5279,29$ & $5264,45$ & $-0,28$ & $(101,101)$\\
26. & $B^0_s$ -- $5366,79$ & $5368,69$ & $+0,04$ & $(102,102)$\\
27. & $B^\pm_c$ -- $6275\pm 1$ & $6296,03$ & $+0,32$ & $(221/2,221/2)$\\
28. & $\chi_{b0}(1P)$ -- $9859,44$ & $9873,03$ & $+0,14$ & $(277/2,277/2)$\\
29. & $\chi_{b0}(2P)$ -- $10232,5$ & $10231,37$ & $-0,006$ & $(141,141)$\\
\hline
\end{tabular}
}
\end{center}
\begin{center}
{\textbf{Table\,II.} Mesons: spin-1 line.}
{\renewcommand{\arraystretch}{1}
\begin{tabular}{|c||l|l|c|l|}\hline
 & State and mass (exp.) -- MeV & Mass (theor.) & Error \% & $(l,\dot{l})$\\ \hline\hline
1. & $\rho(770)$ -- $775,26$ & $777,1$ & $+0,24$ & $(39,38)$\\
2. & $\omega(782)$ -- $782,65$ & $797,16$ & $+1,93$ & $(79/2,77/2)$\\
3. & $K^\ast(892)$ & $901,28$ & $+1,04$ & $(42,41)$\\
4. & $\phi(1020)$ -- $1019,46$ & $1011,78$ & $-0,76$ & $(89/2,89/2)$\\
5. & $h_1(1170)$ & $1177,22$ & $+0,61$ & $(48,47)$\\
6. & $b_1(1235)$ -- $1229,5$ & $1226,78$ & $-0,22$ & $(49,48)$\\
7. & $a_1(1260)$ & $1251,95$ & $-0,63$ & $(99/2,97/2)$\\
8. & $K_1(1270)$ -- $1272\pm 7$ & $1277,37$ & $+0,42$ & $(50,49)$\\
9. & $f_1(1285)$ -- $1281,9$ & $1277,37$ & $-0,35$ & $(50,49)$\\
10. & $\pi_1(1400)$ & $1408,32$ & $+0,59$ & $(105/2,103/2)$\\
11. & $K_1(1400)$ -- $1403\pm 7$ & $1408,32$ & $+0,38$ & $(105/2,103/2)$\\
12. & $K^\ast_1(1410)$ -- $1414\pm 15$ & $1408,32$ & $-0,40$ & $(105/2,103/2)$\\
13. & $f_1(1420)$ -- $1426,4$ & $1435,27$ & $+0,62$ & $(53,52)$\\
14. & $\omega(1420)$ & $1435,27$ & $+0,62$ & $(53,52)$\\
15. & $\rho(1450)$ & $1462,48$ & $+0,86$ & $(107/2,105/2)$\\
16. & $\pi_1(1600)$ & $1602,37$ & $+0,15$ & $(56,55)$\\
17. & $\omega(1650)$ & $1660,11$ & $+0,61$ & $(57,56)$\\
18. & $\phi(1680)$ & $1689,37$ & $+0,56$ & $(115/2,113/2)$\\
19. & $K^\ast(1680)$ & $1689,37$ & $+0,56$ & $(115/2,113/2)$\\
20. & $\rho(1700)$ & $1689,37$ & $-0,62$ & $(115/2,113/2)$\\
21. & $D^\ast(2007)^0$ -- $2006,97$ & $1995,97$ & $-0,55$ & $(125/2,123/2)$\\
22. & $D^\ast(2010)^\pm$ & $1995,97$ & $-0,69$ & $(125/2,123/2)$\\
23. & $\phi(2170)$ -- $2175,97\pm 15$ & $2192,19$ & $+0,79$ & $(131/2,129/2)$\\
24. & $D_1(2420)^0$ & $2432,74$ & $+0,53$ & $(69,68)$\\
25. & $D_{s1}(2460)^\pm$ & $2468,13$ & $+0,33$ & $(139/2,137/2)$\\
26. & $D_{s1}(2536)^\pm$ & $2539,67$ & $+0,14$ & $(141/2,139/2)$\\
27. & $D^\ast_{s1}(2700)^\pm$ -- $2709\pm 4$ & $2722,99$ & $+0,52$ & $(73,72)$\\
28. & $J/\psi(1S)$ -- $3096,916$ & $3108,79$ & $+0,38$ & $(78,77)$\\
29. & $\chi_{c1}(1P)$ -- $3510,66$ & $3520,15$ & $+0,27$ & $(83,82)$\\
30. & $\chi_{c}(1P)$ -- $3525,38$ & $3520,15$ & $-0,16$ & $(83,82)$\\
31. & $\psi(2S)$ -- $3686,09$ & $3691,85$ & $+0,16$ & $(85,84)$\\
32. & $\psi(3770)$ -- $3773,15\pm 0,33$ & $3779,23$ & $+0,16$ & $(86,85)$\\
33. & $\chi(3872)$ -- $3871,69\pm 0,17$ & $3867,63$ & $-0,10$ & $(87,86)$\\
\hline
\end{tabular}
}
\end{center}
\begin{center}
{\renewcommand{\arraystretch}{1.0}
\begin{tabular}{|c||l|l|c|l|}\hline
& State and mass (exp.) -- MeV & Mass (theor.) & Error \% & $(l,\dot{l})$\\ \hline\hline
34. & $\chi(3900)$ -- $3888,7\pm 3,4$ & $3912,22$ & $+0,60$ & $(175/2,173/2)$\\
35. & $\psi(4040)$ -- $4039\pm 1$ & $4047,50$ & $+0,18$ & $(89,88)$\\
36. & $\psi(4160)$ -- $4191\pm 5$ & $4185,09$ & $-0,14$ & $(181/2,179/2)$\\
37. & $\chi(4260)$ -- $4251\pm 9$ & $4278,08$ & $+0,64$ & $(183/2,181/2)$\\
38. & $\chi(4360)$ -- $4354\pm 10$ & $4372,12$ & $+0,42$ & $(185/2,183/2)$\\
39. & $\psi(4415)$ -- $4421\pm 4$ & $4419,51$ & $-0,03$ & $(93,92)$\\
40. & $\chi(4430)$ -- $4478\pm{}^{15}_{18}$ & $4467,12$ & $-0,24$ & $(187/2,185/2)$\\
41. & $\chi(4660)$ -- $4665\pm 10$ & $4660,32$ & $-0,10$ & $(191/2,189/2)$\\
42. & $B^\ast$ -- $5325,1$ & $5316,32$ & $-0,16$ & $(102,101)$\\
43. & $B^\ast_s$ -- $5415,4\pm{}^{1,8}_{1,5}$ & $5421,07$ & $+0,10$ & $(103,102)$\\
44. & $B_1(5721)^0$ & $5741,47$ & $+0,36$ & $(106,105)$\\
45. & $B_{s1}(5830)^0$ -- $5828,4$ & $5850,31$ & $+0,37$ & $(107,106)$\\
46. & $\Upsilon(1S)$ -- $9460,30\pm 0,26$ & $9451,33$ & $-0,09$ & $(136,135)$\\
47. & $\chi_{b1}(1P)$ -- $9892,78$ & $9872,90$ & $-0,20$ & $(139,138)$\\
48. & $\Upsilon(2S)$ -- $10023,26$ & $10015,47$ & $-0,08$ & $(140,139)$\\
49. & $\chi_{b1}(2P)$ -- $10255,46$ & $10231,24$ & $-0,24$ & $(283/2,281/2)$\\
50. & $\Upsilon(3S)$ -- $10355,2$ & $10376,37$ & $+0,20$ & $(285/2,283/2)$\\
51. & $\chi_{b1}(3P)$ -- $10512,1\pm 2,3$ & $10522,51$ & $+0,09$ & $(287/2,285/2)$\\
52. & $\Upsilon(4S)$ -- $10579,4$ & $10595,97$ & $+0,16$ & $(144,143)$\\
53. & $\Upsilon(10860)$ -- $10876\pm 11$ & $10892,35$ & $+0,15$ & $(146,145)$\\
53. & $\Upsilon(11020)$ -- $11019\pm 8$ & $11042,07$ & $+0,21$ & $(147,146)$\\
\hline
\end{tabular}
}
\end{center}
\begin{center}
{\textbf{Table\,III.} Mesons: spin-2 line.}
{\renewcommand{\arraystretch}{1}
\begin{tabular}{|c||l|l|c|l|}\hline
 & State and mass (exp.) -- MeV & Mass (theor.) & Error \% & $(l,\dot{l})$\\ \hline\hline
1. & $f_2(1270)$ -- $1275,5$ & $1276,99$ & $+0,12$ & $(101/2,97/2)$\\
2. & $a_2(1320)$ -- $1318,3$ & $1328,60$ & $+0,78$ & $(103/2,99/2)$\\
3. & $K^\ast_2(1430)$ & $1434,89$ & $+0,34$ & $(107/2,103/2)$\\
4. & $f^\prime_2(1525)$ & $1517,87$ & $-0,50$ & $(55,53)$\\
5. & $\eta_2(1645)$ -- $1617\pm 5$ & $1630,73$ & $+0,85$ & $(57,55)$\\
6. & $\pi_2(1670)$ -- $1672\pm 3$ & $1659,73$ & $-0,73$ & $(115/2,111/2)$\\
7. & $K_2(1770)$ -- $1773\pm 8$ & $1778,28$ & $+0,29$ & $(119/2,115/2)$\\
8. & $K_2(1820)$ -- $1816\pm 13$ & $1808,56$ & $-0,40$ & $(60,58)$\\
9. & $\pi_2(1880)$ & $1869,88$ & $-0,54$ & $(61,59)$\\
10. & $f_2(1950)$ & $1963,77$ & $+0,70$ & $(125/2,121/2)$\\
11. & $f_2(2010)$ & $1995,58$ & $-0,72$ & $(63,61)$\\
12. & $f_2(2300)$ & $2293,37$ & $-0,29$ & $(135/2,131/2)$\\
13. & $f_2(2340)$ & $2327,73$ & $-0,52$ & $(68,66)$\\
14. & $D^\ast_2(2460)^0$ -- $2462,6\pm 0,6$ & $2467,75$ & $+0,20$ & $(70,68)$\\
15. & $D^\ast_2(2460)^\pm$ -- $2464,3\pm 1,6$ & $2467,75$ & $+0,14$ & $(70,68)$\\
16. & $\chi_{c2}(1P)$ -- $3556,2$ & $3562,30$ & $+0,17$ & $(84,82)$\\
17. & $\chi_{c2}(2P)$ -- $3927,2\pm 2,6$ & $3911,83$ & $-0,39$ & $(88,86)$\\
18. & $B^\ast_2(5747)^0$ -- $5739\pm 5$ & $5741,08$ & $+0,04$ & $(213/2,209/2)$\\
19. & $B^\ast_{s2}(5840)^0$ & $5849,92$ & $+0,17$ & $(215/2,211/2)$\\
20. & $\chi_{b2}(1P)$ -- $9912,21$ & $9943,67$ & $+0,32$ & $(140,138)$\\
21. & $\Upsilon(1D)$ -- $10163,7\pm 1,4$ & $10158,68$ & $-0,04$ & $(283/2,279/2)$\\
22. & $\chi_{b2}(2P)$ -- $10268,65$ & $10230,87$ & $-0,37$ & $(142,140)$\\
\hline
\end{tabular}
}
\end{center}
\begin{center}
{\textbf{Table\,IV.} Mesons: spin-3 line.}
{\renewcommand{\arraystretch}{1}
\begin{tabular}{|c||l|l|c|l|}\hline
 & State and mass (exp.) -- MeV & Mass (theor.) & Error \% & $(l,\dot{l})$\\ \hline\hline
1. & $\omega_3(1670)$ -- $1667\pm 4$ & $1659,09$ & $-0,47$ & $(58,55)$\\
2. & $\rho_3(1690)$ -- $1688,8\pm 2,1$ & $1688,34$ & $-0,03$ & $(117/2,111/2)$\\
3. & $K^\ast_3(1780)$ -- $1776\pm 7$ & $1777,64$ & $+0,09$ & $(60,57)$\\
4. & $\phi_3(1850)$ -- $1854\pm 7$ & $1838,45$ & $-0,84$ & $(61,58)$\\
\hline
\end{tabular}
}
\end{center}
\begin{center}
{\textbf{Table\,V.} Mesons: spin-4 line.}
{\renewcommand{\arraystretch}{1}
\begin{tabular}{|c||l|l|c|l|}\hline
 & State and mass (exp.) -- MeV & Mass (theor.) & Error \% & $(l,\dot{l})$\\ \hline\hline
1. & $a_4(2040)$ -- $1995\pm 10$ & $1994,05$ & $-0,05$ & $(64,60)$\\
2. & $f_4(2050)$ -- $2018\pm 11$ & $2026,11$ & $+0,40$ & $(129/2,121/2)$\\
3. & $K^\ast_4(2045)$ -- $2045\pm 9$ & $2058,43$ & $+0,66$ & $(65,61)$\\
\hline
\end{tabular}
}
\end{center}

\section{Baryon sector}
The baryon sector is divided (with respect to a charge) on the two sets of coherent subspaces: charged baryons with half-integer spin (all baryons have half-integer spin) from the subspaces $\bsH^\pm_{\rm phys}(b,0)$; neutral baryons of half-integer spin (states from the coherent subspaces of the type $\bsH^0_{\rm phys}(b,0)$).
\begin{center}
{\textbf{Table\,VI.} Baryons: spin-1/2 line.}
{\renewcommand{\arraystretch}{1}
\begin{tabular}{|c||l|l|c|l|}\hline
 & State and mass (exp.) -- MeV & Mass (theor.) & Error \% & $(l,\dot{l})$\\ \hline\hline
1. & $p$ -- $938,27$ & $933,85$ & $-0,47$ & $(85/2,42)$\\
2. & $n$ -- $939,56$ & $933,85$ & $-0,60$ & $(85/2,42)$\\
3. & $\Lambda$ -- $1115,68$ & $1116,79$ & $+0,09$ & $(93/2,46)$\\
4. & $\Sigma$ -- $1189,37$ & $1189,60$ & $+0,02$ & $(48,95/2)$\\
5. & $\Xi$ -- $1314,86$ & $1316,08$ & $+0,09$ & $(101/2,50)$\\
6. & $\Lambda(1405)$ & $1395,03$ & $-0,70$ & $(52,103/2)$\\
7. & $N(1440)$ & $1448,94$ & $+0,60$ & $(53,105/2)$\\
8. & $N(1535)$ & $1531,72$ & $-0,21$ & $(109/2,54)$\\
9. & $\Lambda(1600)$ & $1588,19$ & $-0,73$ & $(111/2,55)$\\
10. & $\Delta(1620)$ $\approx 1630$ & $1616,80$ & $-0,80$ & $(56,111/2)$\\
11. & $N(1650)$ $\approx 1655$ & $1645,67$ & $-0,56$ & $(113/2,56)$\\
12. & $\Sigma(1660)$ & $1674,80$ & $+0,89$ & $(57,113/2)$\\
13. & $\Lambda(1670)$ & $1674,80$ & $+0,28$ & $(57,113/2)$\\
14. & $N(1710)$ & $1704,18$ & $-0,34$ & $(115/2,57)$\\
15. & $\Sigma(1750)$ & $1733,82$ & $-0,92$ & $(58,115/2)$\\
16. & $\Lambda(1800)$ & $1793,86$ & $-0,34$ & $(59,117/2)$\\
17. & $\Lambda(1810)$ & $1824,27$ & $+0,78$ & $(119/2,59)$\\
18. & $\Delta(1910)$ $\approx 1890$ & $1885,84$ & $-0,22$ & $(121/2,60)$\\
19. & $\Lambda^+_c$ -- $2286,46$ & $2276,76$ & $-0,42$ & $(133/2,66)$\\
20. & $\Sigma_c(2455)$ -- $2453,97\pm 0,14$ & $2450,50$ & $-0,16$ & $(69,137/2)$\\
21. & $\Xi_c$ -- $2467,93\pm{}^{0,28}_{0,40}$ & $2486,01$ & $+0,60$ & $(139/2,69)$\\
\hline
\end{tabular}
}
\end{center}
\begin{center}
{\renewcommand{\arraystretch}{1.0}
\begin{tabular}{|c||l|l|c|l|}\hline
& State and mass (exp.) -- MeV & Mass (theor.) & Error \% & $(l,\dot{l})$\\ \hline\hline
22. & $\Xi^\prime_c(2578)$ & $2557,81$ & $-0,78$ & $(141/2,70)$\\
23. & $\Lambda_c(2595)^+$ -- $2592,25\pm 0,28$ & $2594,09$ & $+0,07$ & $(71,141/2)$\\
24. & $\Omega^0_c(2697)$ & $2704,46$ & $+0,27$ & $(145/2,72)$\\
25. & $\Xi_c(2790)$ & $2779,32$ & $-0,38$ & $(147/2,73)$\\
26. & $\Lambda^0_b$ -- $5620,2\pm 1,6$ & $5660,60$ & $+0,72$ & $(105,209/2)$\\
27. & $\Sigma^+_b$ -- $5807,8\pm 2,7$ & $5823,10$ & $+0,26$ & $(213/2,106)$\\
28. & $\Xi^0_b$ -- $5792,4\pm 3,0$ & $5768,68$ & $-0,40$ & $(106,211/2)$\\
29. & $\Xi^-_b$ -- $5792,4\pm 3,0$ & $5768,68$ & $-0,40$ & $(106,211/2)$\\
30. & $\Sigma^-_b$ -- $5815,2$ & $5823,10$ & $+0,08$ & $(213/2,106)$\\
31. & $\Lambda_b(5912)^0$ -- $5912,11\pm 0,26$ & $5932,71$ & $+0,35$ & $(215/2,107)$\\
32. & $\Xi^\prime_b(5935)^-$ -- $5935,02\pm 0,5$ & $5932,71$ & $-0,04$ & $(215/2,107)$\\
33. & $\Omega^-_b$ -- $6048,0\pm 1,9$ & $6043,34$ & $-0,08$ & $(217/2,108)$\\
\hline
\end{tabular}
}
\end{center}
\begin{center}
{\textbf{Table\,VII.} Baryons: spin-3/2 line.}
{\renewcommand{\arraystretch}{1}
\begin{tabular}{|c||l|l|c|l|}\hline
 & State and mass (exp.) -- MeV & Mass (theor.) & Error \% & $(l,\dot{l})$\\ \hline\hline
1. & $\Delta(1232)$ & $1239,17$ & $+0,58$ & $(99/2,48)$\\
2. & $\Sigma(1385)$ & $1394,77$ & $+0,70$ & $(105/2,51)$\\
3. & $N(1520)$ & $1531,47$ & $+0,75$ & $(55,107/2)$\\
4. & $\Lambda(1520)$ & $1531,47$ & $+0,75$ & $(55,107/2)$\\
5. & $\Xi(1530)$ & $1531,47$ & $+0,09$ & $(55,107/2)$\\
6. & $\Delta(1600)$ & $1587,93$ & $-0,75$ & $(56,109/2)$\\
7. & $\Sigma(1670)$ & $1674,55$ & $+0,27$ & $(115/2,56)$\\
8. & $\Omega^-$ -- $1672,45$ & $1674,55$ & $+0,12$ & $(115/2,56)$\\
9. & $\Lambda(1690)$ & $1703,93$ & $+0,82$ & $(58,113/2)$\\
10. & $N(1700)$ & $1703,93$ & $+0,23$ & $(58,113/2)$\\
11. & $\Delta(1700)$ & $1703,93$ & $+0,23$ & $(58,113/2)$\\
12. & $N(1720)$ & $1733,57$ & $+0,79$ & $(117/2,57)$\\
13. & $\Xi(1820)$ & $1824,01$ & $+0,22$ & $(60,117/2)$\\
14. & $N(1875)$ & $1885,59$ & $+0,56$ & $(61,119/2)$\\
15. & $\Lambda(1890)$ & $1885,59$ & $-0,23$ & $(61,119/2)$\\
16. & $N(1900)$ & $1916,76$ & $+0,88$ & $(123/2,60)$\\
17. & $\Delta(1920)$ & $1916,76$ & $-0,17$ & $(123/2,60)$\\
18. & $\Sigma(1940)$ & $1948,19$ & $+0,42$ & $(62,121/2)$\\
19. & $\Sigma_c(2520)$ -- $2518,41\pm{}^{0,21}_{0,19}$ & $2521,53$ & $+0,12$ & $(141/2,69)$\\
20. & $\Lambda_c(2625)^+$ -- $2628,11\pm 0,19$ & $2630,37$ & $+0,08$ & $(72,141/2)$\\
21. & $\Xi_c(2645)$ & $2630,37$ & $-0,55$ & $(72,141/2)$\\
22. & $\Omega_c(2770)^0$ -- $2765,9\pm 2,0$ & $2779,07$ & $+0,33$ & $(74,145/2)$\\
23. & $\Xi_c(2815)$ -- $2816,6\pm 0,9$ & $2816,89$ & $+0,07$ & $(149/2,73)$\\
24. & $\Sigma^{\ast+}_b$ -- $5829,0\pm 3,4$ & $5822,85$ & $-0,10$ & $(107,211/2)$\\
25. & $\Sigma^{\ast-}_b$ -- $5836,4\pm 2,8$ & $5822,85$ & $-0,23$ & $(107,211/2)$\\
26. & $\Lambda_b(5920)^0$ -- $5919,81\pm 0,23$ & $5932,45$ & $+0,21$ & $(108,213/2)$\\
27. & $\Xi(5945)^0$ -- $5948,9\pm 1,5$ & $5987,64$ & $+0,65$ & $(217/2,107)$\\
28. & $\Xi^\ast_b(5955)^-$ -- $5955,33\pm 0,13$ & $5987,64$ & $+0,54$ & $(217/2,107)$\\
\hline
\end{tabular}
}
\end{center}
\begin{center}
{\textbf{Table\,VIII.} Baryons: spin-5/2 line.}
{\renewcommand{\arraystretch}{1}
\begin{tabular}{|c||l|l|c|l|}\hline
 & State and mass (exp.) -- MeV & Mass (theor.) & Error \% & $(l,\dot{l})$\\ \hline\hline
1. & $N(1675)$ & $1674,04$ & $-0,06$ & $(58,111/2)$\\
2. & $N(1680)$ $\approx 1685$ & $1674,04$ & $-0,35$ & $(58,111/2)$\\
3. & $\Sigma(1775)$ & $1762,95$ & $-0,66$ & $(119/2,57)$\\
4. & $\Lambda(1820)$ & $1823,5$ & $+0,19$ & $(121/2,58)$\\
5. & $\Lambda(1830)$ & $1823,5$ & $-0,35$ & $(121/2,58)$\\
6. & $\Delta(1905)$ $\approx 1880$ & $1885,08$ & $+0,27$ & $(123/2,59)$\\
7. & $\Sigma(1915)$ & $1916,25$ & $+0,06$ & $(62,119/2)$\\
8. & $\Delta(1930)$ $\approx 1950$ & $1947,67$ & $-0,12$ & $(125/2,60)$\\
9. & $\Xi(2030)$ & $2043,49$ & $+0,66$ & $(64,123/2)$\\
10. & $\Lambda(2110)$ & $2108,64$ & $-0,06$ & $(65,125/2)$\\
11. & $\Lambda_c(2880)^+$ -- $2881,53\pm 0,35$ & $2892,77$ & $+0,39$ & $(76,147/2)$\\
\hline
\end{tabular}
}
\end{center}
\begin{center}
{\textbf{Table\,IX.} Baryons: spin-7/2 line.}
{\renewcommand{\arraystretch}{1}
\begin{tabular}{|c||l|l|c|l|}\hline
 & State and mass (exp.) -- MeV & Mass (theor.) & Error \% & $(l,\dot{l})$\\ \hline\hline
1. & $\Delta(1950)$ $\approx 1930$ & $1933,88$ & $+0,20$ & $(125/2,59)$\\
2. & $\Sigma(2030)$ & $2042,72$ & $+0,63$ & $(129/2,61)$\\
3. & $\Lambda(2100)$ & $2107,87$ & $+0,37$ & $(131/2,62)$\\
4. & $N(2190)$ & $2174,05$ & $-0,73$ & $(133/2,63)$\\
\hline
\end{tabular}
}
\end{center}
\begin{center}
{\textbf{Table\,X.} Baryons: spin-9/2 line.}
{\renewcommand{\arraystretch}{1}
\begin{tabular}{|c||l|l|c|l|}\hline
 & State and mass (exp.) -- MeV & Mass (theor.) & Error \% & $(l,\dot{l})$\\ \hline\hline
1. & $N(2220)$ $\approx 2250$ & $2240,22$ & $-0,43$ & $(68,127/2)$\\
2. & $N(2250)$ $\approx 2275$ & $2274,20$ & $+0,03$ & $(137/2,64)$\\
3. & $\Lambda(2350)$ & $2342,93$ & $-0,30$ & $(139/2,65)$\\
\hline
\end{tabular}
}
\end{center}
\begin{center}
{\textbf{Table\,XI.} Baryons: spin-11/2 line.}
{\renewcommand{\arraystretch}{1}
\begin{tabular}{|c||l|l|c|l|}\hline
 & State and mass (exp.) -- MeV & Mass (theor.) & Error \% & $(l,\dot{l})$\\ \hline\hline
1. & $\Delta(2420)$ & $2411,40$ & $-0,35$ & $(71,131/2)$\\
2. & $N(2600)$ & $2590,26$ & $-0,37$ & $(147/2,68)$\\
\hline
\end{tabular}
}
\end{center}
\section{Mass quantization}
The tables I--XI show that masses of elementary particles are proportional to the rest mass of electron $m_e=0,511$ MeV with an accuracy of $0,41\%$. Taking into account mass-energy equivalence principle, one can say that the mass formula
\[
m^{(s)}=m_e\left(l+\frac{1}{2}\right)\left(\dot{l}+\frac{1}{2}\right),
\]
defining the mass (energy) of state (cyclic representation $(l,\dot{l})$), in some sense is equivalent to the well-known relation
\[
E=h\nu,
\]
where the electron mass $m_e$ plays a role of ``mass quantum''.

\end{document}